\documentclass[]{aa}  

\usepackage{scrextend}
\usepackage[colorlinks]{hyperref}

\usepackage{booktabs}
\usepackage{tabularx} 
\let\orgautoref\autoref
\renewcommand{\autoref}
        {\def\equationautorefname{Eq.}%
         \def\figureautorefname{Fig.}%
         \def\sectionautorefname{Sect.}%
         \def\subsectionautorefname{Sect.}%
         \def\subsubsectionautorefname{Sect.}%
         \orgautoref}
         
\usepackage{xcolor}
\definecolor{dark-red}{rgb}{0.9,0.0,0.0}
\definecolor{dark-blue}{rgb}{0.15,0.15,0.9}
\definecolor{dark-green}{rgb}{0.15,0.8,0.15}
\definecolor{medium-blue}{rgb}{0,0,0.9}
\hypersetup{linkcolor={dark-blue},citecolor={dark-blue}, urlcolor={medium-blue}
}

\makeatletter
\renewcommand*\aa@pageof{, page \thepage{} of \pageref*{LastPage}} 
\makeatother

\usepackage{changepage} 
\usepackage{adjustbox}
\usepackage{rotating}

\usepackage{graphicx}
\usepackage{txfonts}
\usepackage{soul}
\usepackage{url}
\usepackage{siunitx}
\usepackage{makecell}
\usepackage[nameinlink,capitalize]{cleveref}
\crefname{section}{Sect.}{Sect.}
\Crefname{section}{Section}{Section}

\begin{document} 

   \title{The CARMENES search for exoplanets around M dwarfs}

    \subtitle{Revisiting the GJ\,581 multi-planetary system with new Doppler measurements from CARMENES, HARPS, and HIRES }
  

\author{A.\,von Stauffenberg\inst{1,2} 
\and T.\,Trifonov\inst{1,3,4} 
\and A.\,Quirrenbach\inst{1}   
\and S.\,Reffert\inst{1}  
\and A.\,Kaminski\inst{1} 
\and S.\,Dreizler\inst{5}  
\and I.\,Ribas\inst{6,7}  
\and A.\,Reiners\inst{5} 
\and M.\,K\"urster\inst{3} 
\and J.\, D.\,Twicken\inst{8,9}  
\and D.\,Rapetti\inst{9,10} 
\and J.\,A.\,Caballero\inst{11} 
\and P.\,J.\,Amado\inst{12} 
\and V.\,J.\,S.~B\'ejar\inst{13,14}  
\and C.\,Cifuentes\inst{11} 
\and S.\,G\'ongora\inst{15} 
\and A.\,P.\,Hatzes\inst{16} 
\and Th.~Henning\inst{3} %
\and D.\,Montes\inst{17} 
\and J.\,C.\,Morales\inst{6,7} 
\and A.\,Schweitzer\inst{18} 
}

         \institute{
         Landessternwarte, Zentrum f\"ur Astronomie der Universt\"at Heidelberg, K\"onigstuhl 12, 69117 Heidelberg, Germany
        \and School of Physics, University College Dublin, Belfield, Dublin 4, Ireland\\
        \email{antonia.vonstauffenberg@ucdconnect.ie}
        \and Max-Planck-Institut f\"ur Astronomie, K\"onigstuhl 17, 69117 Heidelberg, Germany            
        \and
             Department of Astronomy, Faculty of Physics, Sofia University ``St. Kliment Ohridski'', 5 James Bourchier Blvd., BG-1164 Sofia, Bulgaria
         \and Institut f\"ur Astrophysik, Georg-August-Universit\"at, 
              Friedrich-Hund-Platz 1, 37077 G\"ottingen, Germany
        \and Institut de Ci\`encies de l'Espai (ICE, CSIC), Campus UAB, c/ de Can Magrans s/n, 08193 Bellaterra, Barcelona, Spain  
        \and Institut d’Estudis Espacials de Catalunya (IEEC), c/ Gran Capit\`a 2-4, 08034 Barcelona, Spain  
        \and SETI Institute, Mountain View, CA 94043, USA
        \and NASA Ames Research Center, Moffett Field, CA 94035, USA
        \and Research Institute for Advanced Computer Science, Universities Space Research Association, Washington, DC 20024, USA
        \and Centro de Astrobiolog\'{i}a (CSIC-INTA), ESAC campus, 28692 Villanueva de la Ca\~{n}ada, Madrid, Spain 
        \and Instituto de Astrof\'{i}sica de Andaluc\'{i}a (IAA-CSIC), Glorieta de la Astronom\'{i}a s/n, 18008 Granada, Spain
        \and Instituto de Astrof\'isica de Canarias (IAC), 38205 La Laguna, Tenerife, Spain    
        \and Departamento de Astrof\'isica, Universidad de La Laguna (ULL), 38206 La Laguna, Tenerife, Spain
        \and Centro Astron\'onomico Hispano en Andaluc\'ia, Observatorio de Calar Alto, Sierra de los Filabres, 04550 G\'ergal, Almer\'ia, Spain
        \and Th\"uringer Landessternwarte Tautenburg, Sternwarte 5, 07778 Tautenburg, Germany
        \and Departamento de F\'isica de la Tierra y Astrof\'isica and IPARCOS-UCM (Instituto de F\'isica de Part\'iculas y del Cosmos de la UCM), Facultad de Ciencias F\'isicas, Universidad Complutense de Madrid, 28040, Madrid, Spain
        \and Hamburger Sternwarte, Gojenbergsweg 112, 21029 Hamburg, Germany
 }

   \date{Received 29 January 2024 / Accepted 20 March 2024}

 
  \abstract
   {GJ\,581 is a nearby M dwarf known to host a packed multiple planet system composed of two super-Earths and a Neptune-mass planet.
   We present new orbital analyses of the GJ\,581 system, utilizing recent radial velocity (RV) data obtained from the CARMENES spectrograph combined with newly reprocessed archival data from the HARPS and HIRES spectrographs.}
   {Our aim was to analyze the post-discovery spectroscopic data of GJ\,581, which were obtained with CARMENES. In addition, we used publicly available HIRES and HARPS spectroscopic data to seek evidence of the known and disputed exoplanets in this system. We aimed to investigate the stellar activity of GJ\,581 and update the planetary system's orbital parameters using state-of-the-art numerical models and techniques.
   }
   {We performed a periodogram analysis of the available precise CARMENES, HIRES, and HARPS RVs and of stellar activity indicators. We conducted detailed orbital analyses by testing various orbital configurations consistent with the RV data. We studied the posterior probability distribution of the 
   parameters fit to the data and we explored the long-term stability and overall orbital dynamics of 
   the GJ\,581 system.
   }   
   {We refined the orbital parameters of the GJ\,581 system using the most precise and complete set of Doppler data available. Consistent with the existing literature, our analysis confirms that the system is unequivocally composed of only three planets detectable in the present data, dismissing the putative planet GJ\,581\,d as an artifact of stellar activity. Our N-body fit reveals that the system's inclination is $i = 47.0_{-13.0}^{+14.6}$\,deg, which implies that the planets could be up to 30\% more massive than their previously reported minimum masses. Furthermore, we report that the GJ\,581 system exhibits long-term 
   stability, as indicated by the posterior probability distribution, characterized by secular dynamical interactions without the involvement of mean motion resonances.}
   {}


\authorrunning{von Stauffenberg et al.}
\titlerunning{Revisiting the GJ\,581 system}

   \maketitle


\section{Introduction}
\label{sec:Introduction}

As of January 2024, exoplanet 
surveys have yielded over 5,500 confirmed exoplanets. About 1,000 planets have been discovered using the precise radial velocity (RV) method thanks to over four decades of Doppler surveys, whereas more than 3,800 planet candidates 
have been detected using the transit photometry technique, primarily thanks to NASA's highly successful \textit{Kepler} space telescope \citep{Borucki2010} and Transiting Exoplanet Survey Satellite \citep[TESS;][]{Ricker2015}.
The majority of the remaining exoplanet discoveries have been made either via direct imaging \citep[e.g.,][]{Marois2008} or the gravitational microlensing method \citep[e.g.,][]{Gaudi2012}.

Many exoplanets reside in over 870\footnote{Up-to-date list
available on \url{https://exoplanet.eu/}} multiple-planet systems, which are fundamentally important in understanding the formation and evolution of our Solar System and the broader exoplanetary landscape. 
In this context, the precise RV method has exhibited remarkable efficiency in discovering multiple exoplanet systems and characterizing their orbital architectures. The RV data are sensitive to the orbital eccentricity, minimum planetary masses, and in some exceptional cases to the overall dynamical fingerprint of the system, which could allow for the pathways of planetary evolution in the early phases of the protoplanetary disk to be reverse engineered. 

Many multiple-planet systems have been discovered around nearby M dwarfs, which 
are primary targets for a number of planet search surveys via the precise RV method \citep[e.g.,][etc.]{Marcy1998,Delfosse1998,Endl2003,Kurster2003,Butler2006,Zechmeister2009b,Bonfils2013,Reiners2018a}. 
M dwarfs represent a large fraction -- about 72\% -- of the stars in the solar neighborhood \citep{Golovin2023,reyle2021}. Exoplanets induce higher Doppler signals in M dwarfs than in more massive stars such as the Sun. The lower stellar masses allow for the detection of potentially rocky planets in the habitable zone \citep{Kopparapu2013} and multiple-planet systems with relatively packed shorter periods with typical RV semi-amplitudes of a few m\,s$^{-1}$. 
As a result, more than 100 M-dwarf planetary systems have been discovered via the transit and RV methods and half of which are multiple-planet systems. More than 15\% of the known M-dwarf exoplanet systems contain three and more planets, with the record holder being the TRAPPIST-1 system, which is orbited by seven closely packed rocky exoplanets \citep{Gillon2016,Agol2021}. Thanks to TESS, the number of planets around M dwarfs detected with the transit technique and with mass determination from precise RV is constantly growing too \citep[e.g.,][just to mention a few examples of our team]{Trifonov2021Sci,Bluhm2021,Caballero2022,Gonzalez2022,Palle2023}.

M dwarfs, however, pose specific observational challenges due to 
their rather high stellar activity, which can sometimes be mistaken for a 
planetary signal \citep[e.g.,][]{hatzes2016periodic,TalOr2018}. Furthermore, stochastic stellar noise, commonly referred to as RV jitter, typically results in velocity fluctuations on 
the order of 1--2\,m\,s$^{-1}$, for the most inactive stars, a range 
comparable to the Doppler velocity amplitudes induced by low-mass planets. Therefore, securing exoplanet detections in M-dwarf surveys 
typically requires many RV observations with cutting-edge instruments.
 
In recent years, it has become evident that acquiring precise RV measurements in the redder part of the spectrum significantly enhances M-dwarf surveys \citep{Quirrenbach2014,Reiners2018b}. Among the instruments tailored for this purpose is the CARMENES\footnote{Calar Alto high-Resolution search for
M dwarfs with Exo-earths with Near-infrared and optical \'Echelle Spectrographs, \url{http://carmenes.caha.es}} spectrograph \citep{Quirrenbach2016}, which is specifically suited for measurements of M dwarfs. This is due to its operation in the red part of the visible and the near-infrared (NIR) spectral regions, which reduces the influence of stellar activity on the RVs and is often detrimental in studies of these active stars. 

It is essential to continue long-term observing programs of M dwarfs as they have some of the most planet-rich systems (e.g., TRAPPIST-1) and show a close-in habitable zone due to their small size and temperature, which provides an ideal environment for small Earth-like planets. Clarifying the number of planets in these systems, the properties of each planet, and the dynamical relationships between them is only possible with extensive data sets covering long time spans with precise RVs.


\begin{table}[] 
\caption{Stellar parameters of GJ\,581.} 
\label{table:phys_param}    
\centering          
\begin{tabular}{l c r}     
\hline
\hline  
\noalign{\smallskip}        
Parameter & Value & Reference \\  
\noalign{\smallskip}        
\hline    
\noalign{\smallskip}   
Identifiers                                     & BD--07~4003 & Sch86 \\
                                                & Gl 581 & Gli69 \\
                                                & Karmn J15194--077 & Cab16 \\
\noalign{\smallskip}        
\hline    
\noalign{\smallskip}   
$\alpha$ (J2000)                                & 15 19 26.83 & {\em Gaia} DR3 \\ 
$\delta$ (J2000)                                & --07 43 20.2 & {\em Gaia} DR3 \\ 
$\mu_\alpha \cos{\delta}$ [mas\,yr$^{-1}$]      & --1221.278 $\pm$ 0.037 & {\em Gaia} DR3 \\ 
$\mu_\delta$ [mas\,yr$^{-1}$]                   & --97.229 $\pm$ 0.027 & {\em Gaia} DR3 \\ 
$\varpi$ [pc]                                   & 158.718 $\pm$ 0.030 & {\em Gaia} DR3 \\   
$d$ [pc]                                        & 6.3005 $\pm$ 0.0012 & {\em Gaia} DR3 \\ 
$\gamma$ [km\,s$^{-1}$]                         & +9.75 $\pm$ 0.16 & {\em Gaia} DR3 \\ 
$U$ [km\,s$^{-1}$]                              & --25.0660 $\pm$ 0.0084 & This work \\ 
$V$ [km\,s$^{-1}$]                              & --25.6517 $\pm$ 0.0048 & This work \\ 
$W$ [km\,s$^{-1}$]                              & +11.8904 $\pm$ 0.0074 & This work \\ 
\noalign{\smallskip}        
\hline    
\noalign{\smallskip}   
$G^{(a)}$ [mag]                                 & 9.4218 $\pm$ 0.0028 & {\em Gaia} DR3 \\ 
Spectral type                                   & M3.0\,V & Rei95 \\ 
$L_\star$  [$L{_\odot}$]                        & 0.012365 $\pm$ 0.000068 & Cif23 \\
$T_{\mathrm{eff}}$~[K]                          & 3500 $\pm$ 26 & Mar21 \\
$\log g~[\mathrm{\mathrm{cm\,s}}^{-2}]$         & 4.97 $\pm$ 0.11 & Mar21 \\   
{}[Fe/H]                                        & $-$0.08 $\pm$ 0.07 & Mar21 \\ 
$R_\star$    [$R_{\odot}$]                      & 0.302 $\pm$ 0.005 & This work \\
$M_\star$    [$M_{\odot}$]                      & 0.295 $\pm$ 0.010 & This work \\ 
\noalign{\smallskip}        
\hline    
\noalign{\smallskip}   
$P_{\rm rot}$    [$d$]                          & 132.5 $\pm$ 6.3 & Su\'a15 \\
$F_{\rm X}$ [$10^{-14}$\,erg\,s$^{-1}$\,cm$^{-2}$] & 1.8 $\pm$ 0.2 & Loy16\\
$v\sin{i}$    [${\mathrm{km\,s}}^{-1}$]         & $<$ 2.0 & Rei18 \\
pEW(H$\alpha$) [\AA]                            & +0.188 $\pm$ 0.018 & Fuh20 \\
$\log{R'_{\rm HK}}^{(b)}$                       & $-5.48^{+0.17}_{-0.12}$ & This work \\ 
$B$ [G]                                         & 160 $\pm$ 50 & Rei22 \\
Galactic population                             & Young disk& This work \\
\noalign{\smallskip} 
\hline   
\end{tabular}
\tablebib{
    Sch86: \citealt{schonfeld1886}; 
    Gli69: \citealt{gliese1969}; 
    Rei95: \citealt{reid1995}; 
    Su\'a: \citealt{suarez2015}; 
    Cab16: \citealt{Caballero2016};
    Loy16: \citealt{loyd2016};
    Rei18: \citealt{Reiners2018a}; 
    Fuh20: \citealt{fuhrmeister2018}; 
    {\it Gaia} DR3: \citealt{gaia2021}; 
    Mar21: \citealt{marfil2021}; 
    Rei22: \citealt{reiners2022magnetism}; 
    Cif23: \citealt{cifuentes2023}. 
}
\tablefoot{$^{(a)}$ Photometry in passbands different from {\em Gaia}'s $G$ from the blue (Tycho-2's $B_T$) to the mid-infrared (WISE's $W4$) were compiled by \cite{cifuentes2020}. $^{(b)}$ Average $\log{R'_{\rm HK}}$ from 563 FEROS, HARPS, HIRES, and TIGRE measurements compiled by \cite{perdelwitz2021}.}
\end{table}

\begin{table*}
\caption{Published main orbital parameters of all confirmed and conjectural planets in the GJ\,581 system.} 
\label{table:GJ581strikesback}    
\centering          
\begin{tabular}{l cccc l}
\hline
\hline
\noalign{\smallskip}
Planet & $P_{\rm orb}$ {[}d{]}        & $M_{\rm p} \sin{i}$ {[}$M_{\oplus}${]} & $a$ {[}au{]}       & $e$       & Reference         \\ 
\noalign{\smallskip}
\hline
\noalign{\smallskip}
e      & 3.14942 $\pm$ 0.00045  & 1.94  & 0.03 & 0 (fixed)       & \cite{Mayor2009} \\
\noalign{\smallskip}
      & 3.14867 $\pm$ 0.00039  & 1.7 $\pm$ 0.2  & 0.0284533 $\pm$ 0.0000023 & 0 (fixed)       & \cite{Vogt2010} \\
\noalign{\smallskip}
      & 3.14941 $\pm$ 0.00022  & 1.84  & 0.028 & 0 (fixed)       & \cite{Forveille2011} \\
\noalign{\smallskip}
      & 3.1494 $\pm$ 0.0305  & 1.860 $\pm$ 0.406  & 0.028459 $\pm$ 0.000177 & 0 (fixed)       & \cite{vogt2012gj}$^{(a)}$ \\
\noalign{\smallskip}
      & 3.14905 $\pm$ 0.00016  & 1.77 $\pm$ 0.13  & 0.02845621 $\pm$ 0.0000095 & 0.195 $\pm$ 0.073 & \cite{Baluev2013}$^{(b)}$ \\
\noalign{\smallskip}
      & 3.1490 $\pm$ 0.0002  & 1.7 $\pm$ 0.2  & 0.02815 $\pm$ 0.00006 & 0.00 $\pm$ 0.06 & \cite{Robertson2014} \\
\noalign{\smallskip}
      & $3.153^{+0.001}_{-0.006}$ & $1.657^{+0.240}_{-0.161}$  & 0.029 $\pm$ 0.001 & $0.125^{+0.078}_{-0.015}$ & \cite{Trifonov2018a} \\
\noalign{\smallskip}
      & 3.14994$_{-0.00074}^{+0.00074}$  & 1.83$_{-0.10}^{+0.10}$  & 0.02799$_{-0.0003}^{+0.0003}$ & 0.029$_{-0.016}^{+0.019}$       & This work    \\
\noalign{\bigskip}
b      & 5.366$\pm$0.001  & 16.6 & 0.041  & 0 (fixed)       & \cite{Bonfils2005}   \\
\noalign{\smallskip}
      & 5.3683$\pm$0.0003  & 15.7 & 0.041  & 0.02 $\pm$ 0.01       & \cite{Udry2007}   \\
\noalign{\smallskip}
      & 5.36874 $\pm$ 0.00019  & 15.65  & 0.04 & 0 (fixed)       & \cite{Mayor2009}    \\
\noalign{\smallskip}
      & 5.36841 $\pm$ 0.00026  & 15.6 $\pm$ 0.3  & 0.0406163 $\pm$ 0.0000013 & 0 (fixed)       & \cite{Vogt2010} \\
\noalign{\smallskip}
      & 5.36864 $\pm$ 0.00009  & 15.96  & 0.041 & 0 (fixed)       & \cite{Forveille2011} \\
\noalign{\smallskip}
      & 5.3694 $\pm$ 0.0122  & 16.00 $\pm$ 1.17  & 0.0406161 $\pm$ 0.0000609 & 0 (fixed)       & \cite{vogt2012gj}$^{(a)}$ \\
\noalign{\smallskip}
      & 5.368589 $\pm$ 0.000068  & 15.86 $\pm$ 0.16  & 0.04061189 $\pm$ 0.00000034 & 0.022 $\pm$ 0.010 & \cite{Baluev2013}$^{(b)}$ \\
\noalign{\smallskip}
      & 5.3686 $\pm$ 0.0001  & 15.8 $\pm$ 0.3  & 0.04061 $\pm$ 0.00003 & 0.00 $\pm$ 0.03 & \cite{Robertson2014} \\
\noalign{\smallskip}
      & $5.368^{+0.001}_{-0.001}$ & $15.20^{+0.22}_{-0.27}$  & 0.041 $\pm$ 0.001 & $0.022^{+0.027}_{-0.005}$ & \cite{Trifonov2018a} \\
\noalign{\smallskip}
      & 5.36832$_{-0.00010}^{+0.00010}$  & 15.31$_{-0.38}^{+0.38}$ & 0.0399$_{-0.0005}^{+0.0005}$  & 0.02836$_{-0.0087}^{+0.0087}$       & This work   \\
\noalign{\bigskip}
c      & 12.932$\pm$0.007  & 5.03 & 0.073  & 0.16 $\pm$ 0.07       & \cite{Udry2007}   \\
\noalign{\smallskip}
      & 12.9292$\pm$ 0.0047  & 5.36  & 0.07 & 0.17 $\pm$ 0.07       & \cite{Mayor2009} \\
\noalign{\smallskip}
      & 12.9191 $\pm$ 0.0058  & 5.6 $\pm$ 0.3  & 0.072993 $\pm$ 0.000022 & 0 (fixed)       & \cite{Vogt2010} \\
\noalign{\smallskip}
      & 12.9171 $\pm$ 0.0022  & 5.41  & 0.073 & 0 (fixed)       & \cite{Forveille2011} \\
\noalign{\smallskip}
      & 12.9355 $\pm$ 0.0591  & 5.302 $\pm$ 0.881  & 0.072989 $\pm$ 0.000226 & 0 (fixed)       & \cite{vogt2012gj}$^{(a)}$ \\
\noalign{\smallskip}
      & 12.9186 $\pm$ 0.0021  & 5.38 $\pm$ 0.28  & 0.0729286 $\pm$ 0.0000080 & 0.040 $\pm$ 0.044 & \cite{Baluev2013}$^{(b)}$ \\
\noalign{\smallskip}
      & 12.914 $\pm$ 0.002  & 5.5 $\pm$ 0.3  & 0.0721 $\pm$ 0.0003 & 0.00 $\pm$ 0.06 & \cite{Robertson2014} \\
\noalign{\smallskip}
      & $12.919^{+0.003}_{-0.002}$ & $5.652^{+0.386}_{-0.239}$  & 0.074 $\pm$ 0.001 & $0.087^{+0.250}_{-0.016}$ & \cite{Trifonov2018a} \\
\noalign{\smallskip}
      & 12.92015$_{-0.00062}^{+0.00062}$ & 5.01$_{-0.21}^{+0.21}$  & 0.0717$_{-0.0008}^{+0.0008}$  & 0.0161$_{-0.0111}^{+0.0118}$       & This work \\
\noalign{\bigskip}
[g]     & 36.562 $\pm$ 0.052  & 3.1 $\pm$ 0.4  & 0.14601 $\pm$ 0.00014 & 0 (fixed)       & \cite{Vogt2010} \\
\noalign{\smallskip}
      & 32.129 $\pm$ 0.635  & 2.242 $\pm$ 0.644  & 0.13386 $\pm$ 0.00173 & 0 (fixed)       & \cite{vogt2012gj}$^{(a)}$ \\
\noalign{\bigskip}
[d]      & 83.6$\pm$0.7  & 7.7 & 0.25  & 0.20 $\pm$ 0.10       & \cite{Udry2007}   \\
\noalign{\smallskip}
      & 66.80$\pm$ 0.14  & 7.09  & 0.22 & 0.38 $\pm$ 0.09       & \cite{Mayor2009} \\
\noalign{\smallskip}
      & 66.87 $\pm$ 0.13  & 5.6 $\pm$ 0.6  & 0.21847 $\pm$ 0.00028 & 0 (fixed)       & \cite{Vogt2010} \\
\noalign{\smallskip}
      & 66.59 $\pm$ 0.10  & 5.26  & 0.22 & 0 (fixed)       & \cite{Forveille2011} \\
\noalign{\smallskip}
      & 66.671 $\pm$ 0.948  & 5.94 $\pm$ 1.05  & 0.21778 $\pm$ 0.00198 & 0 (fixed)       & \cite{vogt2012gj}$^{(a)}$ \\
\noalign{\bigskip}
[f]     & 433 $\pm$ 13  & 7.0 $\pm$ 1.2  & 0.758 $\pm$ 0.015 & 0 (fixed)       & \cite{Vogt2010} \\
\noalign{\smallskip}
\hline
\end{tabular}
\tablefoot{$^{(a)}$ Astrometric, circular, noninteracting model for a potential five-planet system as preferred by \citet{vogt2012gj}.
$^{(b)}$ Shared-red-jitter model for a potential three-planet system as preferred by \citet{Baluev2013}.}
\end{table*}

In this paper we present a new orbital analysis of the well-known multiple planetary system around the M dwarf GJ\,581, using archival HIRES \citep{Vogt1994}, HARPS \citep{Mayor2003}, and CARMENES data. 
This work presents a detailed extension of our previous work  
\citep{Trifonov2018a}, who studied GJ\,581 together with six other
M-dwarf systems. Here we provide constraints on the orbital inclinations and dynamical masses of the planets, based on a detailed dynamical analysis of the compact three-planet system. In contrast to \citet{Trifonov2018a}, we provide new insights into the nature of the stellar activity of this system as derived from the significant $\sim$ 67\,d signal in the Doppler data. 
To achieve this, we employ a nested sampling (NS) scheme for Bayesian posterior analyses, which combines a self-consistent N-body orbital model and a Gaussian process (GP) model, which serves as a proxy for the stellar activity.
To sum up, the present study represents a more focused and detailed investigation of the orbital and dynamical aspects of GJ\,581, utilizing contemporary Bayesian numerical techniques. 

This work is organized as follows:
In \cref{sec:Literature}, we provide a literature overview of the GJ\,581 multi-planet system and the physical properties of the stellar host, which were estimated elsewhere or in this work. In \cref{sec:Data}, we present the spectroscopic and photometric data used for our orbital characterization and stellar activity analysis. \Cref{sec:Analysis} describes our spectroscopic data analysis methods and results. \Cref{sec:OrbDynAnalysis} presents our orbital analysis results and finally, our summary is given in \cref{sec:ConclusionDiscussion}.

\section{Overview of the \texorpdfstring{GJ\,581}{GJ 581} system}
\label{sec:Literature}
 
GJ\,581 is a very weakly active M3.0\,V star, located just 6.3\,pc away from the Sun. 
Although we use the Gliese-Jahreiss identificator, it was already tabulated in the first catalog of nearby stars of \cite{gliese1969}.
However, it had already been listed as a high proper motion star by many others before, such as \cite{giclas1964}, \cite{Luyten1955}, \cite{porter1930}, \cite{wolf1919},
and \cite{schonfeld1886} in his {\em Bonner Durchmusterung des sudlichen Himmels}.
Having been known for over a century, the star GJ\,581 has been investigated in detail in many works on parallax \citep{adams1926,perryman1997}, photometry \citep{kron1953,leggett1992}, spectroscopy \citep{vyssotsky1946,sousa2008}, chromospheric activity \citep{stauffer1986, wright2004}, multiplicity \citep{fischer1992, warddoug2015}, and many other topics \citep[e.g.,][]{henry1993,delfosse1998b,boyajian2012}. 
We summarise the main astrophysical parameters of GJ\,581 with their references in \cref{table:phys_param}.
We computed the galactocentric space velocities $UVW$ and assigned a galactic population from the {\em Gaia} DR3 astrometry as \cite{montes2001} and the stellar radius and mass from the bolometric luminosity and effective temperature via Stefan-Boltzmann law and the $M_\star$-$R_\star$ relation of \citet{Schweitzer2019}.
Importantly for this work, GJ\,581 has an estimated mass of $0.295 \pm 0.010 \, M_{\odot}$ and a radius of  $0.302 \pm 0.005 \, R_{\odot}$. 
All their activity indicators (long rotational period and slow rotational velocity, faint H$\alpha$, Ca~{\sc ii} H\&K and IRT, ultraviolet, and X-ray emission, and weak magnetic field) point towards an old age inconsistent with the assigned kinematic galactic population.
However, in spite of its old age, GJ\,581 has a debris disk\citep{lestrade2012}.

The GJ\,581 star is also well known because it hosts one of the most extensively discussed and debated multiple-planet systems.
 Different studies have reported varying numbers of planets in this system, ranging from three to six. 
GJ\,581\,b was one of the first exoplanets discovered around an M dwarf star by \citet{Bonfils2005} using HARPS spectra and the RV method.
Two years later,  \citet{udry2007harps} announced the existence of two more planets, namely GJ\,581\,c and d, with further analysis and additional HARPS RVs. 
They confirmed the first planet discovered with an orbital 
period of 5.366\,d and determined orbital  
periods of $P \approx 12.9$ and 83.6\,d for the other two, respectively. They reported minimum masses for the three planets of $\sim 15.7$, 5.0, and 7.7\,$M_{\oplus}$. \citet{Mayor2009} revised the period of GJ\,581\,d to $\sim 66.8$\,d and additionally discovered GJ\,581\,e with the shortest orbital period in the system of 3.15\,d and a minimum mass of $\sim 1.9\,M_{\oplus}$. These values categorized GJ\,581\,b and GJ\,581\,c as Neptune-like planets, while GJ\,581\,d and GJ\,581\,e were classified as super-Earths.

Afterwards, using Keck HIRES \citep[][]{Vogt1994} RVs, \citet{vogt2010lick} suggested the possibility of up to six planets within the system, introducing GJ\,581\,f and GJ\,581\,g with periods of $P \approx$ 433\,d and 37\,d, respectively. This exoplanet discovery announcement generated significant interest within the research field and among the general public, as GJ\,581\,g would have been the first rocky exoplanet positioned within the 
habitable zone of its host star. Subsequently, several follow-up studies conducted an extensive RV analysis of HARPS data available for the system, casting doubt on the existence of planets GJ\,581\,f and GJ\,581\,g \citep[see][]{forveille2011harps}. \citet{vogt2012gj} continued to argue the existence of planet GJ\,581\,g with a follow-up RV analysis and an analysis of dynamical stability, which was further backed by \citet{toth2014dynamical}, who challenged the results of \citet{forveille2011harps}. 

Further, the existence of GJ\,581\,d was questioned by \citet{baluev2013impact}, who analyzed correlated noise components, 
specifically red noise, using data from both HARPS and Keck HIRES. Their 
study confirmed planets GJ\,581\,b, GJ\,581\,c, and GJ\,581\,e but 
suggested that GJ\,581\,d drops to low statistical significance when 
considering the red noise. This was further supported by \citet{Robertson2014} and \citet{hatzes2016periodic}, both of whom 
proposed that the RV signal with a period of $\sim 67$\,d in the RV 
measurements is most likely due to stellar activity rather than an exoplanet in orbit. \citet{Robertson2014} found a significant 
correlation with a Pearson correlation coefficient of $r = -0.31$ between the H$\alpha$ activity index and the HARPS data set after removing the 
dominant signal of GJ\,581\,b. In a similar vein, \citet{hatzes2016periodic} 
showed that H$\alpha$ variations had significant sinusoidal variations that are 180\,deg out of phase with the suggested orbital period of 
GJ\,581\,d, supporting the claim that it is a signal due to stellar activity. 
As a result, the accepted number of planets in this system was (and still is) three, with GJ\,581\,b, GJ\,581\,c, and GJ\,581\,e being confirmed.

The latest update on the orbital parameters of GJ\,581 was presented 
by \citet{Trifonov2018a}, who incorporated the most recent HARPS, HIRES, and CARMENES data sets available then. 
In that work, they assumed the existence of three planets in orbit around GJ\,581, but they did not conduct an in-depth analysis of the stellar activity signal.
The orbital update by \citet{Trifonov2018a} employed a standard $\chi^2$ minimization along with a bootstrap posterior analysis. The stability analysis in that paper, 
however, focused solely on the best-fit solution. 
The main orbital parameters (orbital period $P$, minimum mass $M_{\rm p} \sin{i}$, semi-major axis $a$, and eccentricity $e$) of confirmed GJ\,581\,e, b, and c planets and conjectural GJ\,581\,[g], [d], and [f] planets are displayed, together with the corresponding references, in Table~\ref{table:GJ581strikesback}.

While GJ\,581 is one of the best-studied systems in terms of RV data, none of the planets in the GJ\, 581 system is known to transit so far. 
For example, when it was 
initially discovered by \citet{Mayor2009}, GJ\,581\,e was determined to 
have a 5\% transit probability. \citet{lopez2006limits} were the first to search for transits of the planets orbiting GJ\,581; however, no 
transits were found in their study. 
Afterwards, other teams have also looked for transits of the innermost planet, such as \citet{Forveille2011} and \citet{dragomir2012search}.
The latter 
acquired MOST 
space-based photometry between 2007 and 2009, and found no significant 
candidates. Under the assumption that GJ\,581\,e does indeed produce transits, but with a depth too shallow to be detected, \citet{dragomir2012search} were able 
to rule out a radius larger than 1.62\,$R_{\oplus}$ and thus put lower constraints on its density at $\rho_{\rm e} = 2.5$--3.0\,g\,cm$^{-3}$. Therefore, it is most plausible that the inclinations are not sufficiently close to edge-on to produce transits.

 \begin{figure}
    \centering
    \includegraphics[width=.49\textwidth]{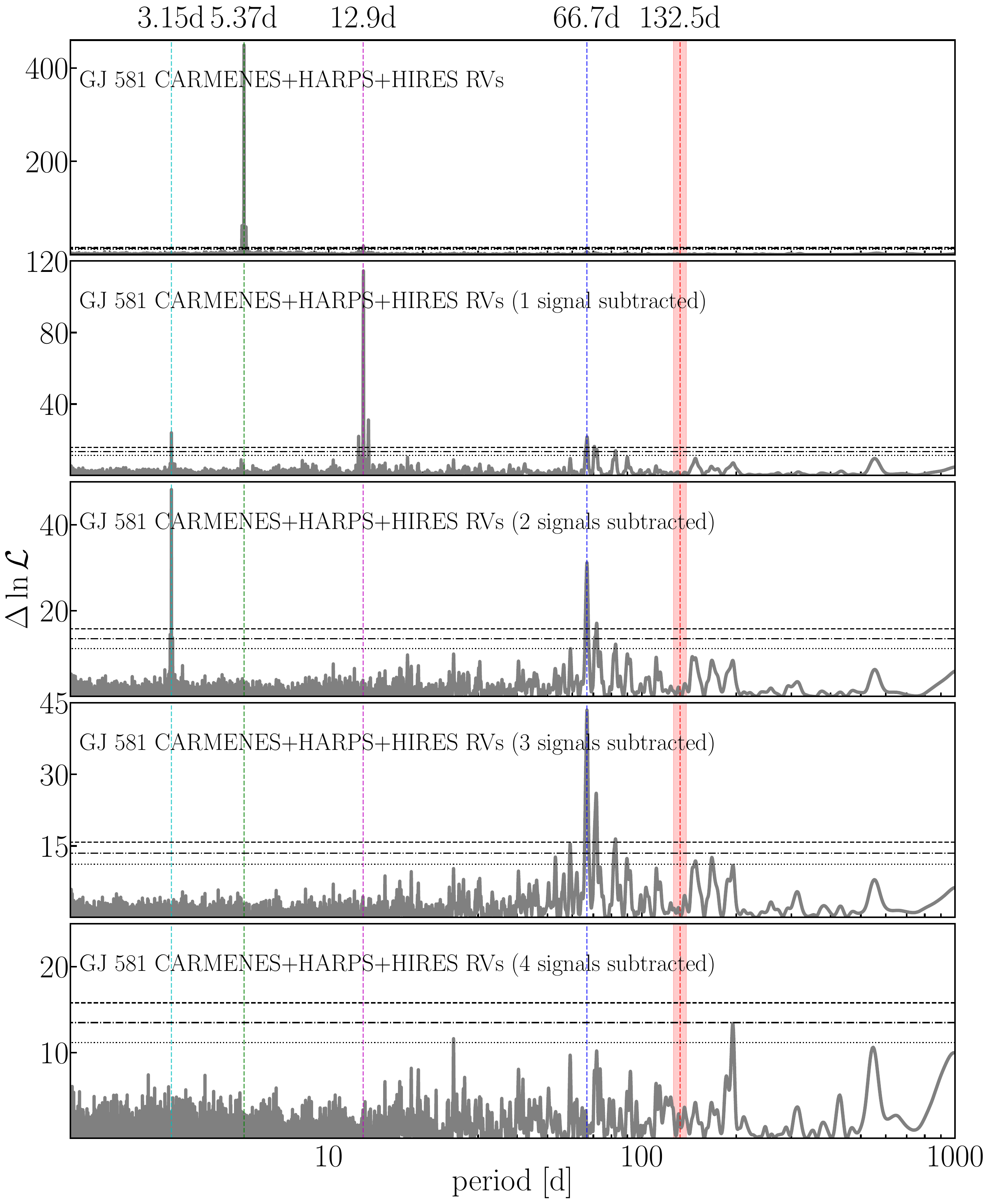}
    \includegraphics[width=.49\textwidth]{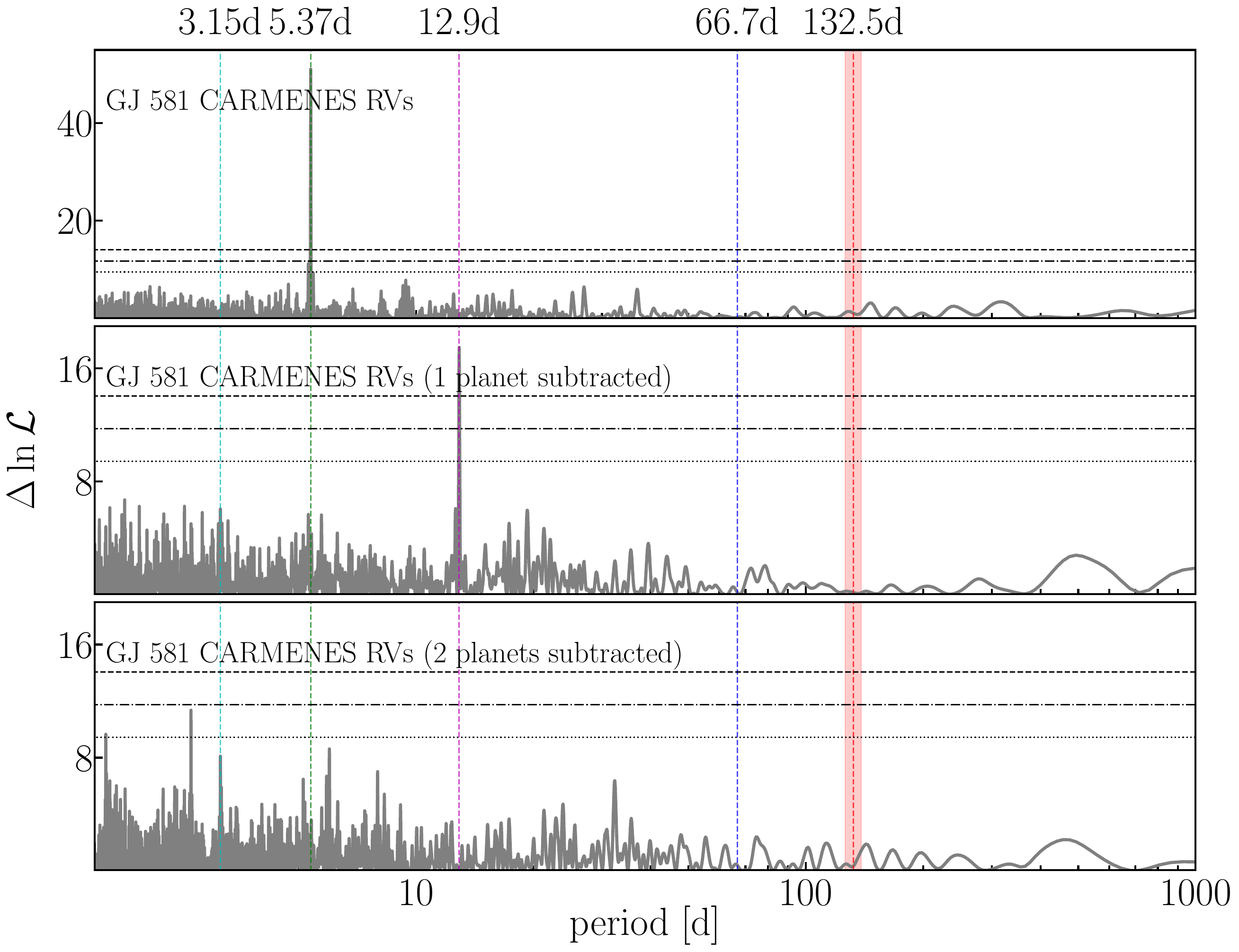}     
    \caption{MLP power spectrum of RV data.
    {\em Top panels}: HIRES, HARPS, and CARMENES. 
    {\em Bottom panels}: CARMENES only.
    Horizontal dashed lines indicate FAP levels of 10\%, 1\%, and 0.1\%, as defined by \citet{Zechmeister2009}. The cyan, green, magenta, and blue vertical lines indicate the orbital periods of GJ\,581\,e, GJ\,581\,b, GJ\,581\,c, and of the conjectural GJ\,581\,[d], respectively.
    The red vertical line represents the rotational period of GJ\,581 ($P_{\rm rot} = 132.5 \pm 6.3$\,d), with shaded regions indicating the associated uncertainty range (see \cref{table:phys_param}). }
    \label{fig:MLP_Rv_periodograms}
\end{figure}


\begin{figure*}[htp]
\centering
\includegraphics[width = 18cm]{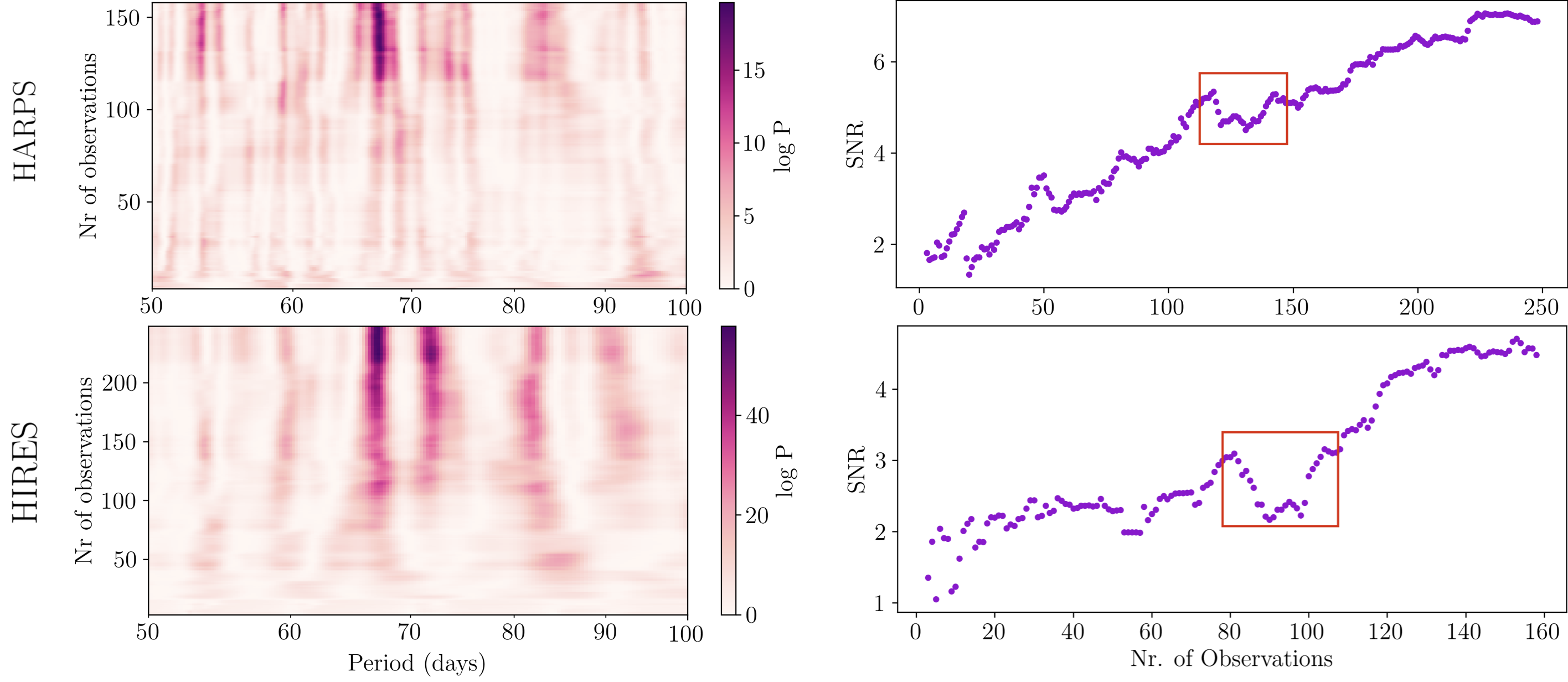}
\caption {S-BGLS (left) and coherence test (right) for HARPS (top) and HIRES (bottom). Red boxes indicate a coinciding dip in significance, which also happens at overlapping time frames.}
\label{fig:SGBLS}
\end{figure*}


\section{Data}
\label{sec:Data}

\subsection{HIRES data}
\label{sec:HIRES}

The HIgh Resolution \'Echelle Spectrograph \citep[HIRES,][]{Vogt1992} is a 
versatile spectrometer at the Keck Observatory, Hawai'i, USA. It 
is installed on the right Nasmyth platform of the 10\,m Keck\,I telescope and operates in the wavelength range of 0.3\,$\mu$m to 1.0\,$\mu$m. In conjunction with a iodine (I$_{\rm 2}$) cell \citep{Marcy1992}, HIRES was the first Doppler machinery capable of delivering relative RV measurements with a precision down to about 3\,m\,s$^{-1}$ \citep{Butler1996}.  Through ongoing instrument and pipeline optimizations, the noise floor has been lowered to around 1\,m\,s$^{-1}$ for bright stars \citep{Butler2017, Luhn2020}, establishing HIRES legacy surveys as a cornerstone in the detection and characterization of small exoplanets.

The HIRES RV measurements of GJ\,581 were collected between August 1999 and August 2014, resulting in a long temporal baseline of 5484\,d. 
We had access to 413 precise RV measurements of GJ\,581 made publicly available through the extensive catalog of 
HIRES spectroscopic RVs and activity indices published by \citet{Butler2017}. This catalog encompasses approximately 65,000 
spectra that were collected over an 18-year period from 1996 to 2014; it includes spectra of GJ\,581 along with those from 1,700 other 
stars. A subsequent reanalysis of the HIRES RV data products by \citet{TalOr2019} led to the adjustment of the data set to account for 
small, yet significant, systematic nightly zero-point variations present 
in the observations. \citet{TalOr2019} showed that the improvement of the 
data was on the order of 1\,m\,s$^{-1}$. 
Besides, a major upgrade was undertaken on the HIRES spectrograph in August 2004, 
which involved the replacement of the CCD \citep{Butler2017}. This upgrade 
impacted the initial 12 RV measurements in the data set; however, no 
significant offset was identified in the data set corrected by 
\citet{TalOr2019}. Consequently, we employed the improved data set published by \citet{TalOr2019} in our 
analyses.

We 
binned the HIRES data into nightly averages, which left us with a set of 158 data points. 
The weighted root-mean-square (wrms) value of the binned data set, being the weight the reported uncertainty of the individual RV, is wrms$_{\rm HIRES} = 9.46$\,m\,s$^{-1}$, which indicates the scatter or variability of the obtained RVs. 
The median RV uncertainty of the binned data set, denoted as $\hat\sigma_{\rm HIRES}$, is 1.76\,m\,s$^{-1}$.

\begin{figure*}
\centering
\includegraphics[width=18cm]{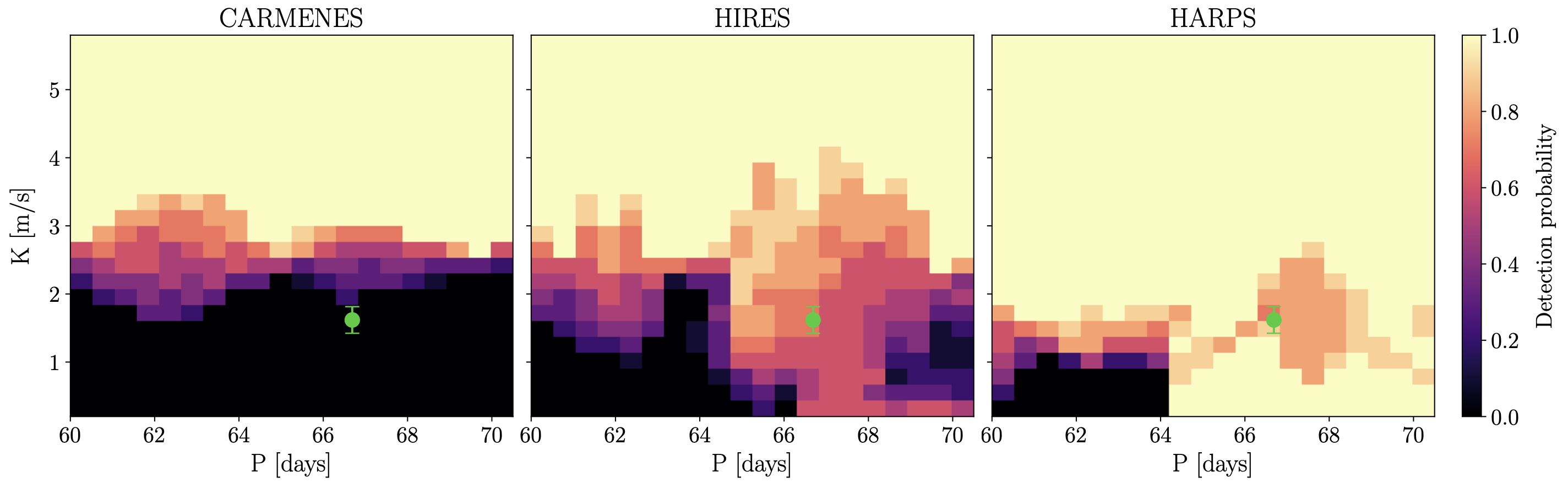}
\caption{Injection-retrieval detection probability grids for CARMENES, HIRES, and HARPS, from left to right.
Higher detection probabilities are marked with lighter colors.
The location in the semi-amplitude-orbital period ($K$-$P$) diagram of the conjectural planet GJ\,581\,[d] is marked by a green dot.}
\label{fig:Imshow plot}
\end{figure*}

\subsection{HARPS data}
\label{sec:HARPS}
 
The HARPS spectrograph \citep{Mayor2003} is operated at the ESO 3.6\,m Telescope at the La Silla Observatory, Chile and is one of the most powerful RV instruments by delivering 
precision below 1 m\,s$^{-1}$. We utilized the {\tt serval} \citep{Zechmeister2018} pipeline for precise RV extraction.
We obtained time series of several stellar activity 
indicators, such as the H$\alpha$ line, the sodium D lines, the chromatic index (CRX), and the differential line width 
(dLW). These indicators are fundamentally important for characterizing and 
quantifying stellar activity of GJ\,581 and can give insights on how 
different types of stellar activity influence and affect RV measurements\citep{Zechmeister2018}.
We also inspected the data products from the official ESO-HARPS Data 
Reduction Software (DRS) pipeline, from which we collected additional 
activity indicators related to the spectral line profiles, which may also 
show evidence of stellar activity. The DRS time series that we utilized 
were the full-width half maximum (FWHM), the contrast (CON), and the CCF 
bisector slope (BIS), which offer information about spectral line 
profile variations and magnetic activity \citep{Queloz2001}.

We retrieved a total of 250 high-precision RV data points of GJ\,581 obtained between May 2004 and May 2012 with a total temporal baseline of 2909\,d. Nightly binning of the data set slightly reduced the RV data set to 248 data points, which have a wrms$_{\rm HARPS} =  9.68$\,m\,s$^{-1}$ and a median RV uncertainty of $\hat\sigma_{\rm HARPS} = 1.05$\,m\,s$^{-1}$.
The HARPS RV measurements, their corresponding activity index data, as well as their respective uncertainties are tabulated in \cref{table:HARPS_1}.

\subsection{CARMENES data}
\label{sec:CARMENES}

GJ\,581 was part of the regular monitoring program involving over 350 M-dwarf targets within the CARMENES guaranteed time observation program (see \citealt{Reiners2018a} and \citealt{Ribas2023} for details). 
Within the CARMENES survey, we obtained 54 pairs of optical (VIS, 0.52\,$\mu$m to 0.96\,$\mu$m) and near-infrared (NIR, 0.96\,$\mu$m to 1.71\,$\mu$m) spectra of GJ\,581. These spectra represent the most 
recent data in the study, with the first data points acquired in January 
2016 and the last in July 2019, covering a time span of 1265\,d. The 
typical exposure time was approximately 20\,min, with the objective of 
achieving an signal-to-noise ratio (S/N) of 150 in the $J$ band.

All spectra underwent standard CARMENES data processing following the 
procedure detailed by \citet{Caballero2016}. These calibrated spectra were subsequently processed using {\tt serval} to compute precise RVs in both the VIS and NIR ranges. For an M3.0\,V star such as GJ\,581, the CARMENES NIR precision is typically poorer, as demonstrated by \citet{Reiners2018b} and \citet{bauer2020}, since the  Doppler information contained in the spectral lines is lower than in the optical channel. Consequently, we did not use the CARMENES NIR RV data for updating the orbital information of GJ\,581. As a result, when referring to CARMENES RVs throughout our work, we specifically mean using the CARMENES VIS data. Furthermore, we harnessed the extensive CARMENES precise RV data set collected during the survey period to compute minor nightly zero-point (NZP) systematics, making corrections to enhance the precision, as thoroughly explained by \citet{TalOr2019} and \citet{Trifonov2018a, Trifonov2020}.

The CARMENES RV data have a median error of $\hat\sigma_{\rm CARMENES} = 1.29$\,m\,s$^{-1}$ and a weighted-root-mean-square value of wrms$_{\rm CARMENES} = 9.21$\,m\,s$^{-1}$. 
Concurrently with the RV extraction from the CARMENES spectra, {\tt serval} generated time series data for various stellar activity indices, such as CRX, dLW, calcium infrared triplet 
(Ca~{\sc ii} IRT $\equiv$ CaIRTa,b,c), H$\alpha$, Na\,D$_1$, and Na\,D$_2$ 
\citep{Zechmeister2018, Schoefer2019}. Additionally, using the {\tt raccoon} pipeline \citep{Lafarga2020}, we computed the FWHM, BIS, and CON from the CARMENES spectra. 
The CARMENES data, which include the RV data set and activity indices along with their respective uncertainties, are available in \cref{table:CARM_MAIN}.

\begin{table*}[htp]
\caption{Comparison between GJ\,581 models described in the text.
} 
\label{table:stat_param}    
\centering          
\begin{tabular}{ l l r r r r r r}     
\hline
\hline  
\noalign{\smallskip}
No. of & Model   \hspace{60mm} & $\ln{\mathcal{L}}$   &  BIC &  $\ln{\mathcal{Z}}$  &    $\Delta\ln\mathcal{L}$   &    $\Delta$BIC &  $\Delta\ln{\mathcal{Z}}$ \\ 
planets \\
\noalign{\smallskip}
\hline    
\noalign{\smallskip}
0 & No planet (Base model)                          & $-$1691.96         & 3420.63 &    $-$1708.58  & 0 & 0  & 0\\ 

  1 & GJ\,581\,b                         &     $-$1239.84      & 2547.12 &  $-$1278.62 & 452.12 & 873.51 &  429.96  \\ 
  
  2 & GJ\,581\,b + c                         &     $-$1120.95     & 2340.01  &  $-$1177.59 & 571.01 & 1080.61 & 530.99 \\ 
  
  3 & GJ\,581\,b + c + e                        &     $-$1070.43      & 2269.62 &  $-$1139.26  & 621.53 & 1151.00  & 569.32 \\ 

  4 & GJ\,581\,b + c + e + GP                         &     $-$993.60      & 2115.96 & $-$1081.08 & 698.36 & 1304.66 & 627.50 \\
  
  5 & GJ\,581\,b + c + e + GP (N-body, coplanar edge-on)                          &    $-$988.38     & 2105.51 & $-$1083.88 & 703.58 &  1315.11 & 624.70 \\ 
  6 & GJ\,581\,b + c + e + GP (N-body, coplanar inclined)                         &    $-$986.88    & 2108.65 & $-$1077.76 & 705.08 &  1311.98 & 630.82\\ 
\noalign{\smallskip}
\hline   
\end{tabular}
\end{table*}

\subsection{TESS data}
 \label{sec:TESS}

We investigated the TESS data containing precise photometry for GJ\,581. TESS observed GJ\,581 (TIC\,36853511) in sector 51 between 22 April 2022 and 18 May 2022. GJ\,581 is not expected to be observed again by the end of \textit{TESS} cycle 6, but is expected to be observed by \textit{TESS} in sector 91 of cycle 7 (Spring 2025).
The sector 51 data, obtained in 2-min cadence integrations, were retrieved from the Mikulski Archive for Space 
Telescopes\footnote{\url{https://mast.stsci.edu}}.
For this target, the Science Processing Operations Center \citep[SPOC;][]{SPOC} at NASA Ames Research Center provided both simple aperture photometry (SAP) and systematics-corrected photometry derived from the {\em Kepler} Presearch Data Conditioning algorithm \citep[PDCSAP,][]{Smith2012, Stumpe12}. PDCSAP light curves undergo correction for contamination from nearby stars and instrumental systematics stemming from pointing drifts, focus changes, and thermal transients. As described in \cref{sec:Transit}, in the SPOC pipeline, a portion of the sector 51 light curve was excluded due to scattered light from Earth and the Moon.

\section{Data analysis and results}
\label{sec:Analysis}

\subsection{Signal analysis tools}
\label{Sec4.1}
 
To perform data and orbital analyses of the GJ\,581 system, we employed the versatile {\tt Exo-Striker} exoplanet 
toolbox\footnote{\url{https://github.com/3fon3fonov/exostriker}} \citep{Trifonov2019_es, Trifonov2020}. This open-source Python library offers an intuitive graphical user interface and integrates various public 
tools for exoplanet analysis. Notable features include a generalized Lomb-Scargle periodogram \citep[GLS,][]{Zechmeister2009} and a maximum $\ln\mathcal{L}$  
periodogram \citep[MLP,][]{Baluev2008, Zechmeister2019}. Additionally, it incorporates the {\tt wotan} package for transit photometry detrending \citep{Hippke2019} and a 
transit period search via the {\tt transitleastsquares} package 
\citep[TLS,][]{Hippke2019b}, among others. We employed these tools for 
signal analysis in this study. Beyond {\tt Exo-Striker}, we also employed 
Stacked Bayesian Generalized Lomb-Scargle periodograms \citep[S-BGLS][]{Mortier2015, Mortier2017} for a more detailed period analysis of the RV 
data. The subsequent subsections elaborate on specifics regarding our 
data analysis setup.

 \subsection{Periodogram analysis}
 \label{sec:RV Periodogram}

For the periodogram analysis of the RVs and activity index data of GJ\,581, we employed MLP. In 
contrast to the GLS periodogram, the MLP does not optimize the least-squares for each periodogram frequency \citep[see][]{Zechmeister2009}, 
but it optimizes the $\ln\mathcal{L}$ metric and quantifies its power as $\Delta\ln\mathcal{L}$ relative to the null (flat) model for each 
frequency. As a result, the MLP allows for flexible parameterization 
during the frequency scan and can accommodate additive mutual data 
offsets and data variance (i.e., the ``jitter'' estimate added in 
quadrature to the uncertainty budget), which can be applied to each 
tested data set individually. While the computation of the MLP is more 
resource-intensive than the GLS periodogram, it has been proven more 
suitable for conducting a rigorous period search in combined data time series.

The upper part of \Cref{fig:MLP_Rv_periodograms} illustrates the MLP power spectrum of the combined RV data. The top panel shows the $\ln\mathcal{L}$ accounting only for offsets and jitter. 
Subsequent MLP panels demonstrate the results obtained with a pre-whitening approach, which involves the successive subtraction of the most prominent period from the data via the best-fitting Keplerian model. 
The last panel displays the MLP of the residuals after fitting for all potential planetary (and stellar activity) signals.
The abscissa represents the periods at which these signals occur on a logarithmic scale, while the ordinate axis quantifies the signal significance using the 
$\Delta\ln{\mathcal{L}}$ metric. 

The first panel of \cref{fig:MLP_Rv_periodograms} shows that the combined RV data exhibit a significant 5.37\,d signal (green vertical line), consistent with the initial detection of GJ\,581\,b by \citet{Bonfils2005}. 
Following the removal of this dominant signal using a Keplerian model, the second panel reveals a robust signal corresponding to the confirmed 12.9\,d period planet GJ\,581\,c (magenta vertical line). Moreover, a signal surpassing the FAP threshold emerges at 3.15\,d, consistent with the innermost planet, GJ\,581\,e (cyan vertical line). Additionally, a 66.7\,d signal, originally attributed to a planet by \citet{Mayor2009}, becomes evident.
Considering the 12.9\,d signal as a planetary candidate and constructing a simultaneous two-planet model, the 3.15\,d and 66.7\,d period signals gain significance in the residuals. The 3.15\,d signal prevails in strength, leading to its adoption as the subsequent third planet.
 
In the fourth panel of \cref{fig:MLP_Rv_periodograms}, the 66.7\,d signal retains its significance, prompting its inclusion in the pre-whitening as a Keplerian signal. The subsequent fifth panel depicts the residuals, where no significant signals are discernible. The lack of signal within the fifth panel supports the inference that the 66.7\,d signal is not an alias of a distinct signal. Further analysis is warranted to determine the nature of the latter signal, whether planetary or stemming from stellar activity.

\begin{figure}[]
\centering
\includegraphics[width=.45\textwidth]{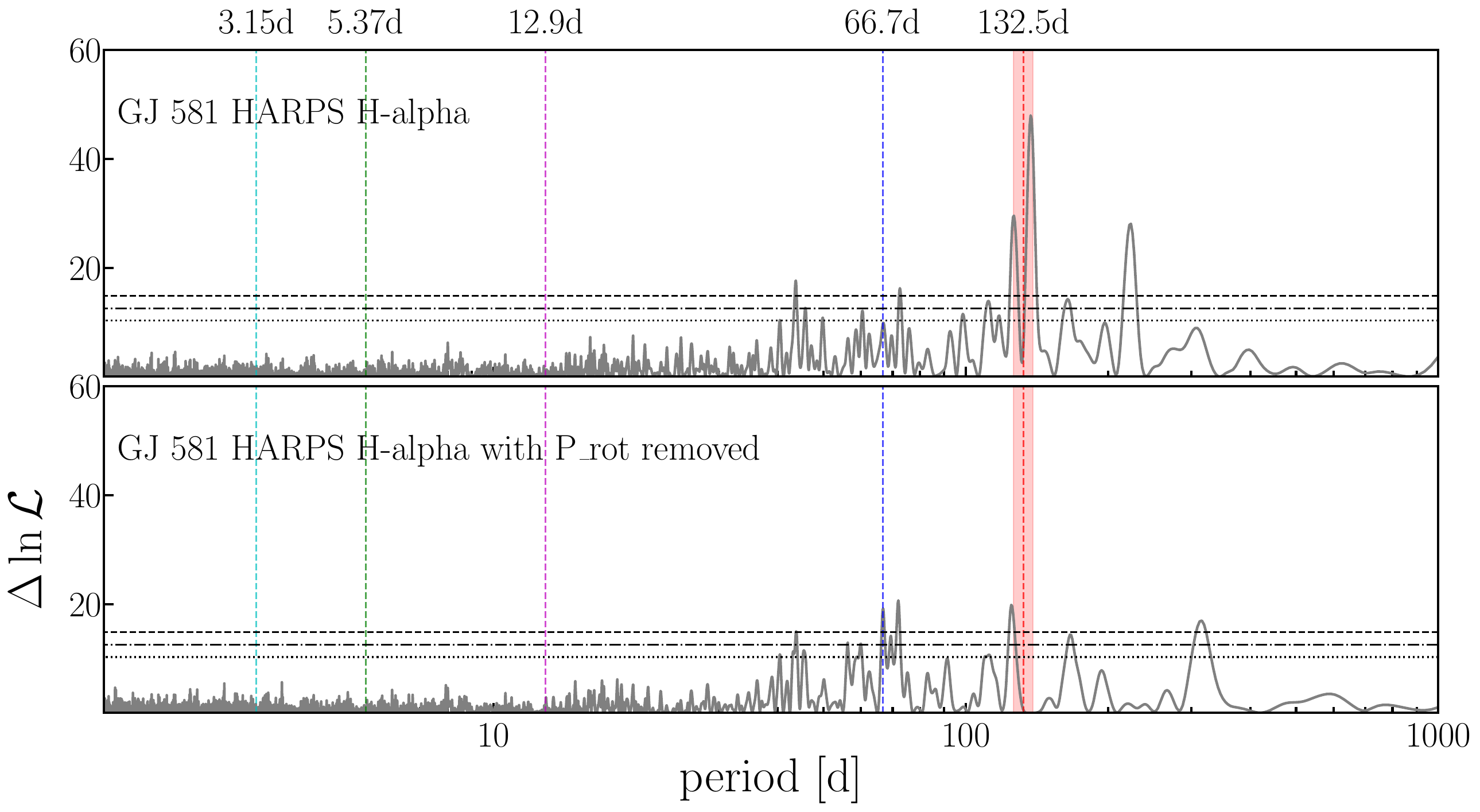}
\caption{MLP power spectrum of the HARPS H$\alpha$ index. {\em Top panel}: Base index with no alterations. {\em Bottom panel}: Residuals after the main periodic signal associated with the stellar rotation was removed.}
\label{fig: Halpha_HARPS}
\end{figure}

We also inspected the individual RV data sets with an MLP to seek
for mutual signal consistency. The MLP cascade of the HARPS data revealed the presence of three planets, GJ\,581\,b, c, and e, along with a potential stellar activity signal at 66.7\,d. MLP analysis of the HIRES data alone successfully reproduced the GJ\,581\,b and c planets, as well as the 66.7\,d signal. However, the precision of the data is likely insufficient to detect the lowest RV amplitude planet GJ\,581\,e. The MLP results for our CARMENES data are presented in the lower part of \cref{fig:MLP_Rv_periodograms}. After pre-whitening, no significant signals were identified following the removal of the second planet with a period of $\sim$ 12.9\,d. Nevertheless, the residuals of the two-planet model revealed an insignificant signal at 3.15\,d, suggesting consistency between the CARMENES and HARPS data. With a larger number of RVs, CARMENES would likely have detected GJ\,581\,e. Notably, no signal appears at a period of 66.7\,d in the CARMENES MLP. This lack hints at the possibility that the 66.7\,d signal may be indicative of stellar activity, more prominent in the blue (where HARPS and HIRES operate) and less detectable or absent in the redder spectral region probed by CARMENES. However, it is also plausible that the CARMENES time series is not sensitive to this signal, an aspect explored in \cref{sec: Injection Retrieval}.

 \subsection{Activity analysis}
 \label{sec:Activity Periodogram}

\Cref{Fig:Fullperiodogram} presents the MLP power spectra of all the RV time series used in this study, from CARMENES, HARPS, and HIRES, along with those of the corresponding spectroscopic activity indices. The HARPS H$\alpha$ series exhibits significant activity signals within the uncertainty range of the stellar rotational period at $132.5 \pm 6.3$\,d. The Na\,D$_1$ and Na\,D$_2$ time series also show strong power close to that range.
The 66.7\,d signal, at half the rotational period, could be the second harmonic. Alternatively, its lower significance could result from the overshadowing strength of the 132.5\,d signal and, therefore, a strong activity signal independent of the rotational period of the star. To explore this further, we removed the 132.5\,d signal in the H$\alpha$ periodogram by fitting a sine model. As depicted in \cref{fig: Halpha_HARPS}, the residuals reveal a significant signal at the 66.7\,d period, along with other significant signals at periods of 43.7\,d, 45.8\,d, 71.7\,d, and 316.8\,d. 
The increase in the 66.7\,d signal thus suggests an independent activity signal strong enough to appear in the RV data, undermining the existence of the conjectural planet GJ\,581\,[d]. However, caution is warranted, as modeling the stellar activity signal near the rotational period with a sinusoid may not be ideal due to potential variations in the semi-amplitude ($K$) or phase of stellar activity over time. 
We attempted a similar analysis with both the Na\,D$_1$ and Na\,D$_2$ time series, but without any clear-cut outcome, as expected due to the strong signal not precisely aligning with the stellar rotational period. 

\begin{figure}
\centering
\includegraphics[width=.45\textwidth]{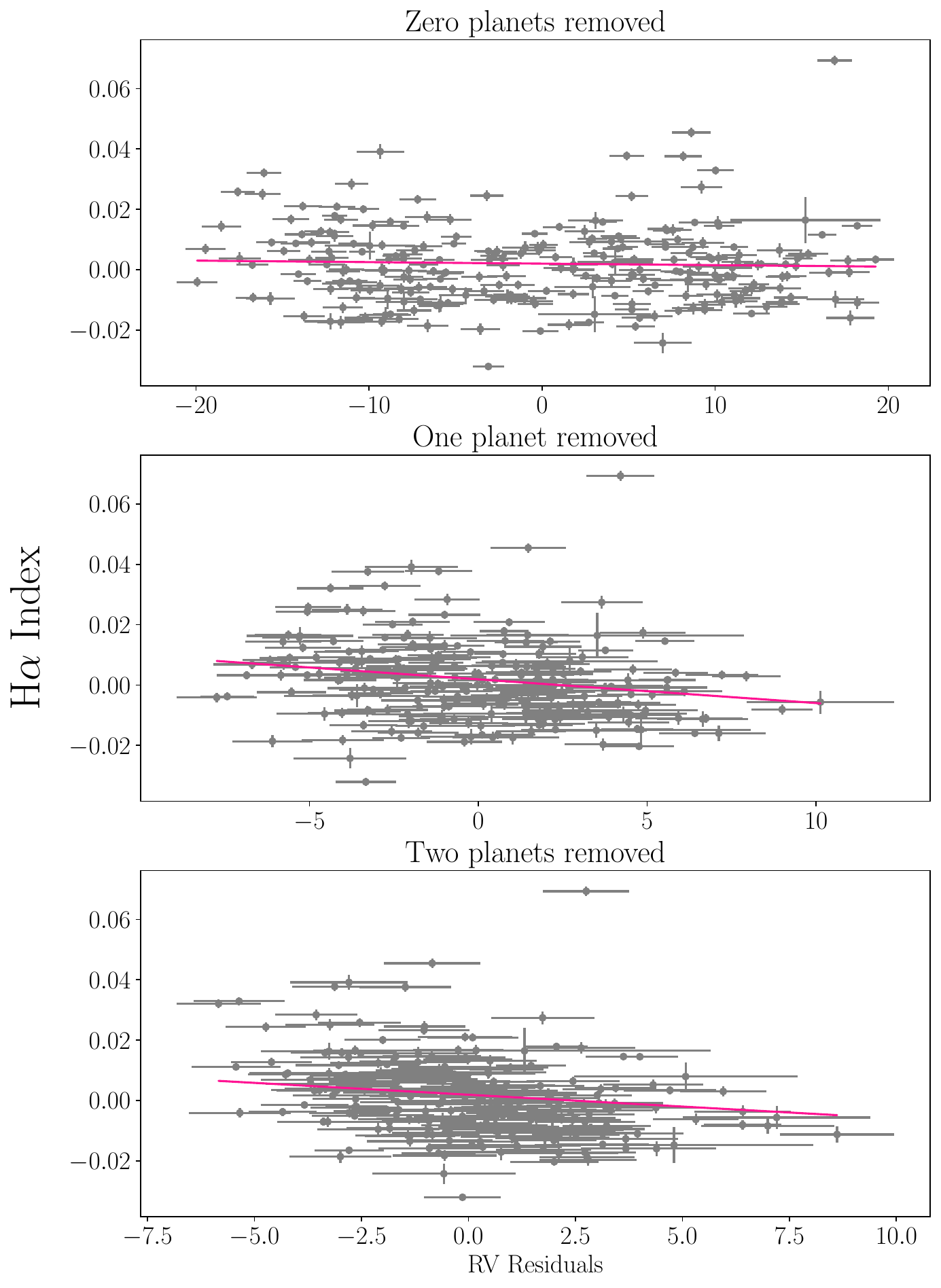}
 \caption{Correlation plots between the HARPS RV residuals and H$\alpha$ activity time series. The first panel shows the correlation when no planet is fit, the second when the 5.36\,d is removed and the third panel when both the 5.35\,d and 12.9\,d planets are removed.}
\label{fig:correlation plot}
\end{figure}

\begin{table*}[ht]
\centering
\caption{{Parameter estimates of the Gl\,581 system from joint N-body modeling of RVs from CARMENES, HIRES, HARPS, and including a GP model for stellar activity. Listed are the best-fit parameters and 1-$\sigma$ uncertainties from the NS posterior analysis.}}    \label{table:NS_post}   
\begin{tabular}{lrrrrrrrr}     
\hline
\hline  
\noalign{\smallskip}
& & Median and 1-$\sigma$ & & &
Maximum $-\ln\mathcal{L}$ &    \\  
\noalign{\smallskip}
\hline 
\noalign{\smallskip}
Parameter \hspace{0.0 mm}& Planet e & Planet b & Planet c& \hspace{10.0 mm}Planet e & Planet b & Planet c  \\
\noalign{\smallskip}
\hline 
\noalign{\smallskip}
$K$ [m\,s$^{-1}$]             &           1.8$_{-0.1}^{+0.1}$ &          12.6$_{-0.1}^{+0.1}$ &           3.1$_{-0.1}^{+0.1}$ & 1.9 & 12.6 & 3.1\\ \noalign{\smallskip}
$P$ [d]                     &           3.1499$_{-0.0001}^{+0.0001}$ &           5.3683$_{-0.0001}^{+0.0001}$ &          12.9201$_{-0.0006}^{+0.0006}$ &3.151 &5.368 & 12.920\\ 
\noalign{\smallskip}
$e$                           &           0.029$_{-0.016}^{+0.019}$ &           0.0284$_{-0.009}^{+0.008}$ &           0.0161$_{-0.0111}^{+0.0118}$ & 0.025& 0.038& 0.017 &\\ 
\noalign{\smallskip}
$\omega$ [deg]                &         235.4$_{-39.8}^{+23.7}$ &          70.4$_{-14.6}^{+16.5}$ &         226.4$_{-103.8}^{+80.9}$ & 243.4 & 71.1 & 237.0 & \\ \noalign{\smallskip}
$MA$ [deg]             &          31.5$_{-21.4}^{+34.5}$ &         214.2$_{-16.9}^{+14.6}$ &         156.9$_{-80.3}^{+103.3}$ & 35.8 &212.9 &146.8 &\\ \noalign{\smallskip}
$a$ [au]                      &           0.02799$_{-0.0003}^{+0.0003}$ &           0.0399$_{-0.0005}^{+0.0005}$ &           0.0717$_{-0.0008}^{+0.0008}$ & 0.028 & 0.040 & 0.072 &\\ 
\noalign{\smallskip}
$m_p$ [$M_{\oplus}$]   &           1.83$_{-0.10}^{+0.10}$ &           15.31$_{-0.38}^{+0.38}$ &           5.01$_{-0.21}^{+0.21}$ & 1.89 & 15.24 & 5.03 & \\ \noalign{\smallskip}
\hline 
\noalign{\smallskip}
GP$_{\rm Rot}$ amp. [m\,s$^{-1}$]  &  &        15.1$_{-3.7}^{+5.3}$ &            &             &  10.4 &   & & \\ 
\noalign{\smallskip}
GP$_{\rm Rot}$ time scale [d]  & &    706.4$_{-81.3}^{+247.8}$ &           &             & 400.1 &   &   & \\ 
\noalign{\smallskip}
GP$_{\rm Rot}$ period [d]  & &          66.8$_{-2.1}^{+1.5}$ &             &             & 66.2 &   &   & \\ 
\noalign{\smallskip}
GP$_{\rm Rot}$ fact.  & &         0.19$_{-0.06}^{+0.09}$ &            &             & 0.21 &   &   & \\ 
\noalign{\smallskip}

RV$_{\rm off}$ CARM. [m\,s$^{-1}$]& &         $-1.4_{-1.6}^{+1.7}$ & & & $-$1.9 &\\ \noalign{\vskip 0.9mm}
RV$_{\rm off}$ HIRES [m\,s$^{-1}$]& &         $-0.3_{-1.3}^{+1.1}$ & & & $-$0.3 & \\ \noalign{\vskip 0.9mm}
RV$_{\rm off}$ HARPS [m\,s$^{-1}$]& &          0.4$_{-1.3}^{+1.3}$  & & & 0.4& \\ \noalign{\vskip 0.9mm}
RV$_{\rm jit}$ CARM. [m\,s$^{-1}$]& &          0.9$_{-0.4}^{+0.4}$  & & & 0.9 & \\ \noalign{\vskip 0.9mm}
RV$_{\rm jit}$ HIRES [m\,s$^{-1}$]& &          1.5$_{-0.2}^{+0.2}$  & & & 1.5 & \\ \noalign{\vskip 0.9mm}
RV$_{\rm jit}$ HARPS [m\,s$^{-1}$]& &          0.6$_{-0.1}^{+0.1}$  & & & 0.6 & \\ 
\noalign{\smallskip}
\hline 
\end{tabular}
\end{table*}

\begin{table*}[ht]
\centering
\caption{{Same as \cref{table:NS_post}, but for the coplanar inclined system.}}   
\label{table:NS_post2}   
\begin{tabular}{lrrrrrrrr}     
\hline
\hline  
\noalign{\smallskip}
& & Median and 1-$\sigma$ & & &
Maximum $-\ln\mathcal{L}$ &    \\  
\noalign{\smallskip}
\hline 
\noalign{\smallskip}
Parameter \hspace{0.0 mm}& Planet e & Planet b & Planet c& \hspace{10.0 mm}Planet e & Planet b & Planet c  \\
\noalign{\smallskip}
\hline 
\noalign{\smallskip}
$K$ [m\,s$^{-1}$]             &           1.8$_{-0.1}^{+0.1}$ &          12.3$_{-0.1}^{+0.1}$ &           3.1$_{-0.1}^{+0.1}$ & 1.8 & 12.6 & 3.1\\ \noalign{\vskip 0.9mm}
$P$ [d]                     &           3.1481$_{-0.0004}^{+0.0004}$ &           5.3686$_{-0.0001}^{+0.0001}$ &          12.9211$_{-0.0007}^{+0.0008}$ &3.149 &5.368 & 12.921\\ \noalign{\vskip 0.9mm}
$e$                           &           0.012$_{-0.008}^{+0.015}$ &           0.0342$_{-0.010}^{+0.009}$ &           0.032$_{-0.021}^{+0.027}$ & 0.015& 0.030& 0.001 &\\ \noalign{\vskip 0.9mm}
$\omega$ [deg]                &         226$_{-55}^{+91}$ &          54$_{-14}^{+13}$ &         16$_{-89}^{+61}$ & 193.6& 43.6 & 84.1 & \\ \noalign{\vskip 0.9mm}
$MA$ [deg]             &          101$_{-92}^{+58}$ &         231$_{-13}^{+14}$ &         7$_{-59}^{+89}$ & 52.7 &241.1& 303.3 &\\ \noalign{\vskip 0.9mm}
$a$ [au]                      &           0.02799$_{-0.0003}^{+0.0003}$ &           0.0399$_{-0.0005}^{+0.0005}$ &           0.0718$_{-0.0009}^{+0.0008}$ & 0.028 & 0.040 & 0.072 &\\ \noalign{\vskip 0.9mm}
$m_p$ [$M_{\oplus}$]   &           2.48$_{-0.42}^{+0.70}$ &           20.5$_{-3.5}^{+6.2}$ &           6.81$_{-1.16}^{+0.21}$ & 2.33 & 19.17 & 6.29 & \\ \noalign{\vskip 0.9mm}
$i$ [deg]                     &          47$_{-13}^{+15}$ &       47$_{-13}^{+15}$   &          47$_{-13}^{+15}$ & 52.7& & & \\ \noalign{\vskip 0.9mm}\noalign{\smallskip}
\hline 
\noalign{\smallskip}
GP$_{\rm Rot}$ amp. [m\,s$^{-1}$]  & &         11.9$_{-2.4}^{+3.1}$ &            &             &  9.0 &   & & \\ \noalign{\vskip 0.9mm}
GP$_{\rm Rot}$ time scale [d]  & &    517.1$_{-89.6}^{+317.8}$ &           &             & 407.0 &   &   & \\ \noalign{\vskip 0.9mm}
GP$_{\rm Rot}$ period [d]  & &          66.4$_{-0.8}^{+1.0}$ &             &             & 66.6 &   &   & \\ \noalign{\vskip 0.9mm}
GP$_{\rm Rot}$ fact.  & &         0.32$_{-0.17}^{+0.11}$ &            &             & 0.05 &   &   & \\ \noalign{\vskip 0.9mm}

RV$_{\rm off}$ CARM. [m\,s$^{-1}$]& &         $-1.7_{-1.4}^{+1.5}$ & & & $-$1.4 &\\ \noalign{\vskip 0.9mm}
RV$_{\rm off}$ HIRES [m\,s$^{-1}$]& &        $-0.4_{-1.1}^{+1.1}$ & & & 0.3 & \\ \noalign{\vskip 0.9mm}
RV$_{\rm off}$ HARPS [m\,s$^{-1}$]& &          0.5$_{-1.1}^{+1.1}$  & & & 0.7& \\ \noalign{\vskip 0.9mm}
RV$_{\rm jit}$ CARM. [m\,s$^{-1}$]& &          1.6$_{-0.3}^{+0.3}$  & & & 0.9 & \\ \noalign{\vskip 0.9mm}
RV$_{\rm jit}$ HIRES [m\,s$^{-1}$]& &          1.7$_{-0.2}^{+0.2}$  & & & 1.6 & \\ \noalign{\vskip 0.9mm}
RV$_{\rm jit}$ HARPS [m\,s$^{-1}$]& &          1.0$_{-0.1}^{+0.1}$  & & & 0.7 & \\ \noalign{\smallskip}

\hline 
\end{tabular}
\end{table*}

\begin{figure*}
    \includegraphics[width=9cm]{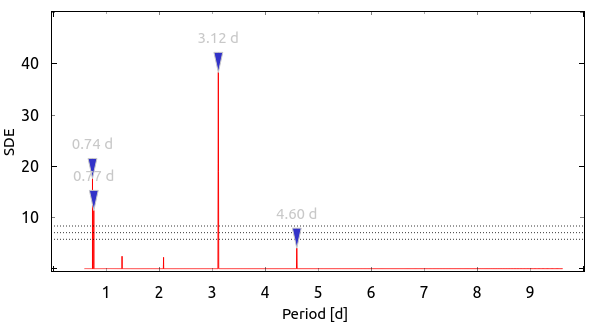} \put(-70,110){TLS}
        \includegraphics[width=9cm]{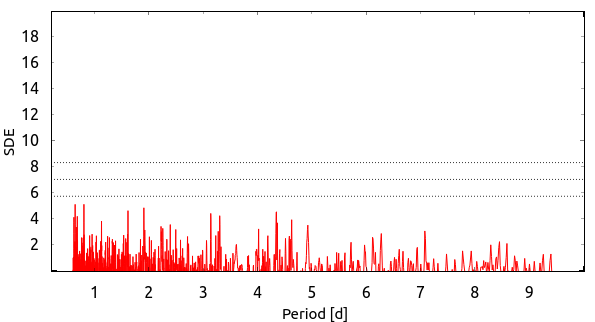} \put(-70,110){TLS (o-c)}\\
    \includegraphics[width=9cm]{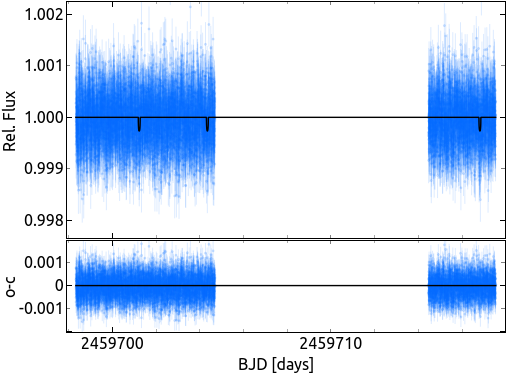}
     \includegraphics[width=9cm]{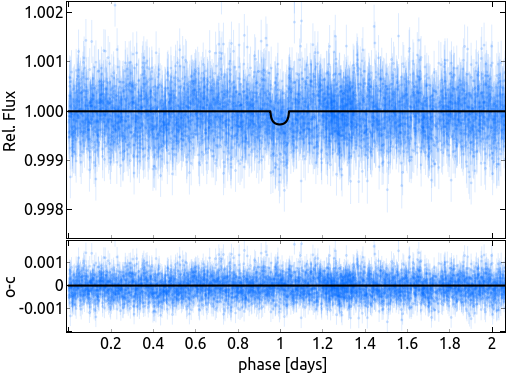}    
    \caption{{\em Top left panel:} TLS spectrum of the publicly available PDCSAP detrended light curves from Sector 51. These data exhibit significant TLS power with a periodicity of 3.12\,d, suggesting a potential transit detection with a prominent proximity to the orbital period of the RV-confirmed planet GJ\,581\,e, with a period of 3.15\,d. {\em Top right panel:} TLS spectrum residuals with no signal left after a Keplerian model is fit.
    {\em Bottom left panel:} Keplerian transit model to the {\sc TESS} data of GJ\,581. {\em Bottom right panel:} Phase-folded representation of the model to the data. While these data can be effectively fit with a Keplerian model,
    further more detailed reanalysis of the {\sc TESS} data from sector 51 pointed out that the 3.12\,d is a marginal false-positive detection and is not related to GJ\,581\,e (see text for details).
    }
    \label{Fig:transt_plots}
\end{figure*} 
 
Additional evidence for the nature of the signal being activity-based is given in \cref{fig:correlation plot}, which shows the correlation between the pre-whitened RVs and the H$\alpha$ index.
As expected, there is almost no correlation in panel 1, as the 5.36\,d (b) planet dominates the RV variations. In panels 2 and 3, when first the 5.36\,d planet and then the 12.9\,d (c) planet are removed, the correlation becomes stronger, with Pearson's correlation coefficients of $-$0.22 and $-$0.33, respectively. These correlation coefficients agree with the results of \citet{Robertson2014}, who found a correlation of $-$0.31 after removing the 5.36\,d planet and declared the GJ\,581\,d planet a false positive on this basis. A similarly strong correlation is present between the HIRES RV data and the Ca\,{\sc II} $S$ index. Here Pearson's correlation coefficients are +0.29 and +0.33 (for one and two planets removed, respectively), further solidifying that there is a correlation between the RV residuals and H$\alpha$ index. 
No significant correlation was found for other activity data. 
In reference to the study by \citet{Robertson2014}, \citet{anglada2015comment} highlighted that correlations after a pre-whitening process might include contributions from actual signals. While acknowledging this concern, we point out that it is not applicable here, as we do not make any quantitative inferences, but only illustrate the likely scenario of stellar activity mimicking a planetary signal.

To further investigate the 66.7\,d signal and to determine its origin, we used S-BGLS periodograms and coherence testing. While similar analyses were performed by \citet{Mortier2017} using the available HARPS data, we inspected the signal in more depth with data from both HARPS and HIRES.
\Cref{fig:SGBLS} shows the results from our S-BGLS  analysis and a signal coherence test, in which the RVs are consecutively included and
the significance S/N$_{K}$ 
at the specific period under investigation (66.8\,d)  
is plotted against the number of data points. For both HARPS and HIRES, the 66.7\,d signal increases in significance overall, which was expected due to the strong periodogram peak. However, there are distinct variations in the log-probability of the signal for both instruments. If it were to be a planet, we would expect the detection significance to rise continuously, which is not the case for either HARPS \citep[as also pointed out by][]{Mortier2017} or HIRES. The dips within the two plots bear a close resemblance to each other (red squares), 
and they span coinciding time frames from late 2008 to early 2010. The coinciding time frame suggests a ``quieter'' period of activity for GJ\,581 and strengthens the conclusion that the origin of the signal at 66.7\,d is stellar activity.

\subsection{Injection-retrieval map analysis of the \texorpdfstring{66.7\,d}{66.7 d} signal}
 \label{sec: Injection Retrieval}
 
We generated injection-retrieval maps for each instrument by simulating the injection of a Keplerian signal in the RV data sets. 
We took residuals from our best-fit model (see below), 
free of any signal and injected sinusoids with a range of periods ($P$), semi-amplitudes ($K$), and mean anomalies ($MA$). 
This process effectively introduced a synthetic planet-like signal into the RV data. 
The signal was subsequently retrieved by generating a GLS periodogram for each parameter combination and by determining whether a planetary signal could be detected with FAP of 0.1\% for a given period range. 
Therefore, we were able to analyse the individual RV data sets and could quantify their capability to detect significant signals at specific periods and semi-amplitudes of the RV curve, particularly in proximity to the signal of the conjectural planet GJ\,581\,d. 
 
We were mainly interested in the CARMENES map, as we wanted to test 
whether CARMENES could detect the 66.7\,d RV signal. The 
first panel of \cref{fig:Imshow plot} shows a very narrow transition zone between the signal being detectable or not. The signal, whose position is signified by a green marker, has a semi-amplitude too low for being detectable in the CARMENES data at a significance of 
FAP of 0.1\%. In contrast, the HARPS data have a high detection 
probability between 70\% and 90\% and the HIRES data a moderate detection probability of around 70\%. 
Thus, we concluded that the number of RV data points from CARMENES (54) is insufficient to detect the potential 66.7\,d planet. This outcome is consistent with the results from HIRES and HARPS, as both of these data sets are larger even after binning the data, by factors of $\sim 5 \times$ (249 data points, HARPS) and $\sim 3 \times$ (159 data points, HIRES).

In addition, the lack of a signal due to activity in the CARMENES RVs is also to be expected as the wavelength range covered by this instrument minimizes the effect of stellar activity in spectroscopic data, due to the RVs being derived from a redder range of the visible spectrum in comparison to HARPS. This could explain the complete absence of any peaks near 66.7\,d in the CARMENES data (see the bottom panel of \cref{fig:MLP_Rv_periodograms}). 
We note that while the possible rotation period is prominently visible in  HARPS H-alpha data, it is conspicuously absent in the periodograms generated from CARMENES activity 
indices. Therefore, it is not unlikely that GJ\,581 was experiencing a quiet phase during the CARMENES observational baseline, thus, potentially contributing to the nondetection of the $\sim$66.7\,d signal.


\subsection{Transit search in the TESS data}
 \label{sec:Transit}

We retrieved the raw PDCSAP TESS light curve for transit search.
The PDCSAP data set is by default corrected for dominant systematics, but small systematics are often evident in the light curve. We applied our own normalization scheme using the {\tt wotan} transit photometry detrending package, which is conveniently wrapped in the {\tt Exo-Striker}.  We used a spline and a robust (iterative) squared-exponential GP kernel for capturing the nonperiodic variation of the light curve \citep[][]{Hippke2019}. 
We thus derived a normalized, flat TESS light curve, which we used in our further orbital analysis.

We carried out a transit signal search on the TESS light curve using the TLS algorithm \citep{Hippke2019b}.
Remarkably, the TLS power spectrum reveals compelling evidence of a shallow, transit-like signal with a periodicity of 3.12\,d. The signal detection efficiency power level is 36.5, by far surpassing the recommended detection threshold of $>8$ as suggested by \citet{Aigrain2016}.
However, we found only three transit-like occurrences with mid-transit epochs of  BJD = 2459701.229, 2459704.346, and 2459716.811. 
These three events are at the bare minimum to claim transit detection, but the signal is intriguing since it is very close to the orbital period of the RV-confirmed planet GJ\,581\,e, with a period of 3.15\,d. \Cref{Fig:transt_plots} shows the TLS power spectrum and a follow-up transit model fit to the TESS light curve.

Our initial hypothesis asserted that gravitational interactions between the planets in the GJ\,581 system, over the combined 
observational baseline (RV + TESS), could lead to 
orbital period oscillations of GJ\,581\,e, resulting in a somewhat 
shorter period of 3.12\,d at the epoch of TESS sector 
51. Such a shift would imply relatively strong transit 
timing variations (TTVs). Utilizing the {\tt ttvfaster} code 
\citep{Agol2016}, we calculated analytic TTV predictions for a range of dynamical masses and eccentricities of the GJ\,581 three-planet system. Our analysis indicates that GJ\,581\,e could 
exhibit TTV amplitudes of only a few minutes and not the $\sim 30$\,min observed. Moreover, a transit-only fit suggests very 
shallow transit depths, corresponding to a planet radius of 
$\sim 0.6\,R_\oplus$, which combined with the RV-estimated 
minimum mass of GJ\,581\,e would result in a nonphysical bulk 
planetary density. Our confidence in the reality of the transit-like signal was further diminished when we were not able to find consistent results even when considering the possibility of a grazing transit.

Nevertheless, we conducted an extensive RV and transit photo-dynamical modeling with posterior analysis using the {\tt Exo-Striker} and the {\tt flexi-fit}\footnote{\url{https://gitlab.gwdg.de/sdreizl/exoplanet-flexi-fit}} codes; we were unable to find a common model that fits the three transit events and the available RV data. 
We note in passing that we explored fits involving other intriguing possibilities. 
For instance, we observed that the 3.12\,d period is the 1-year alias of the 3.15\,d signal 
seen in the RV data. Additionally, significant signals at periods of 0.74\,d  and 0.77\,d  
were identified in \Cref{Fig:transt_plots}. These periods are close to the 1-day alias of 
3.12\,d and 3.15\,d, which is approximately 0.76\,d. Consequently, we inferred that the true period of GJ\,581\,e could be 3.12\,d. Therefore, we considered the true period of GJ\,581\,e 
to be $\sim$ 3.12\,d or $\sim$ 0.76\,d. We conducted separate NS parameter 
scans, imposing strong priors at these orbital periods. Unfortunately, these models did not 
converge well when employing multi-Keplerian or photo-dynamical schemes.

To dispel any doubts, as a final test, we reprocessed the TESS photometry using a bespoke, optimized pixel level decorrelation \cite[PLD;][]{Luger_2016}-based light curve (see Rapetti et al.~in prep. for further details), which recovered segments of the data that were initially excluded by the SPOC pipeline due to scattered light. However, this custom PLD-based light curve did not contain a significant 3.12\,d or $\sim$ 0.76\,d signals. Therefore, we concluded that the 3.12\,d signal is very likely a marginal false-positive and did not discuss it further in our work.

\begin{figure*}
\centering
\includegraphics[width=\textwidth]{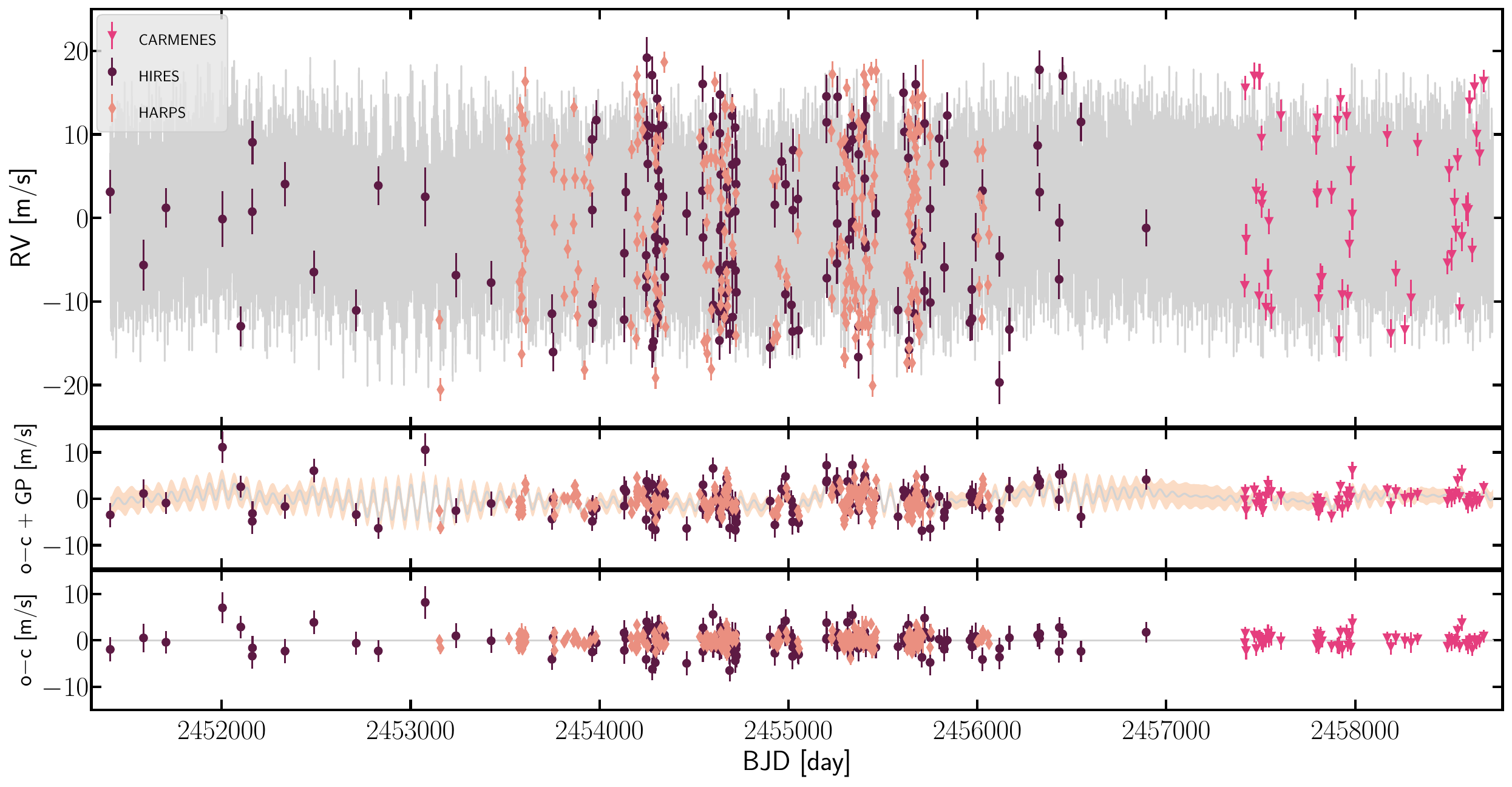}\\
\includegraphics[width=5.95cm]{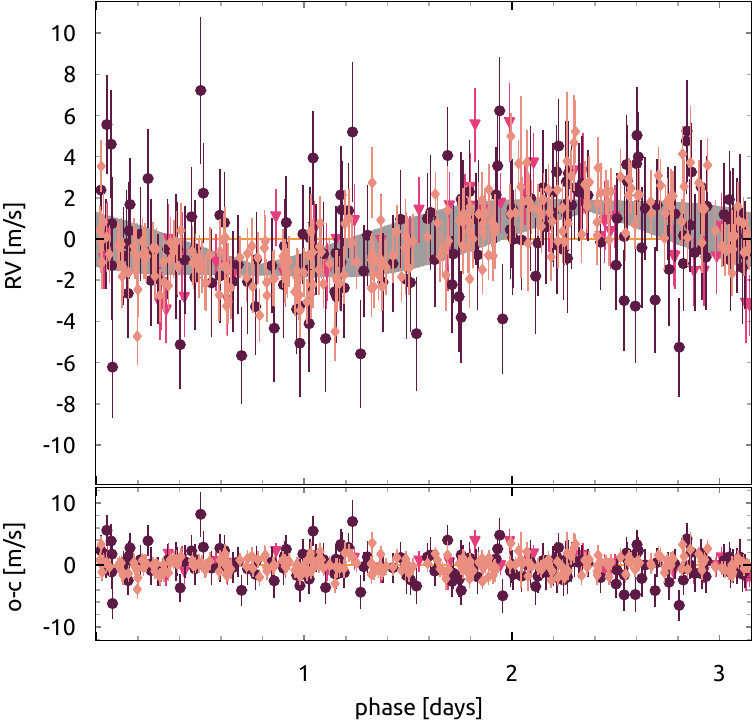} \put(-40,150){GJ\,581\,e}
\includegraphics[width=5.95cm]{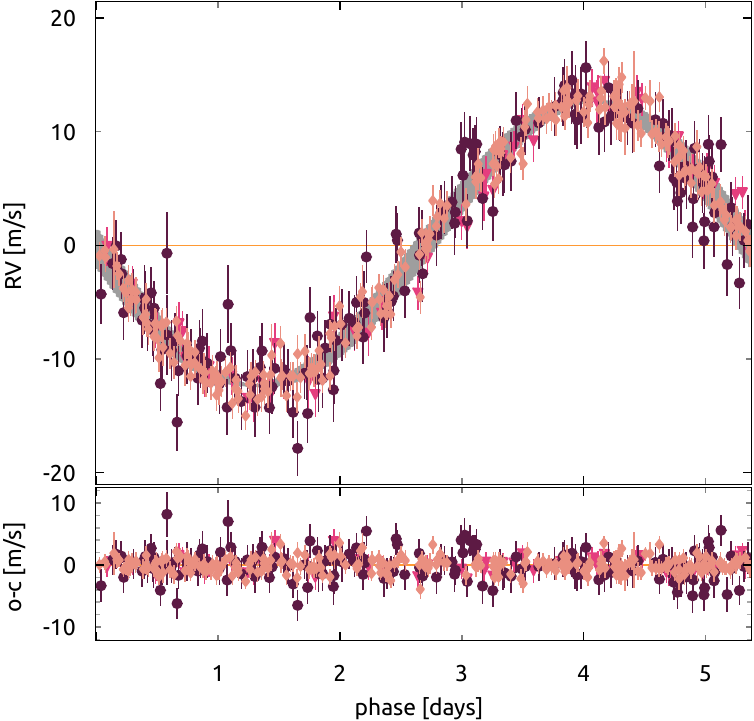} \put(-140,150){GJ\,581\,b}
\includegraphics[width=5.95cm]{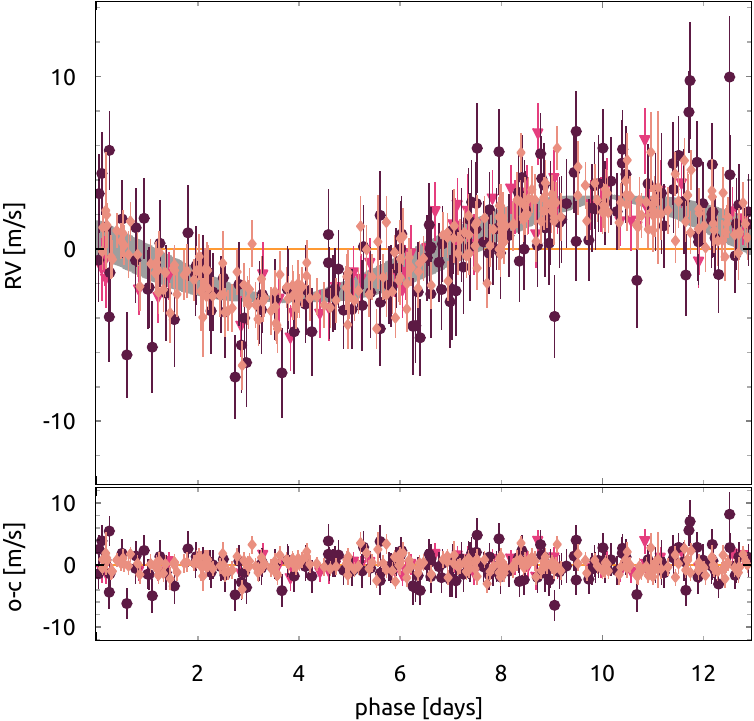} \put(-140,150){GJ\,581\,c}\\
\includegraphics[width=\textwidth]{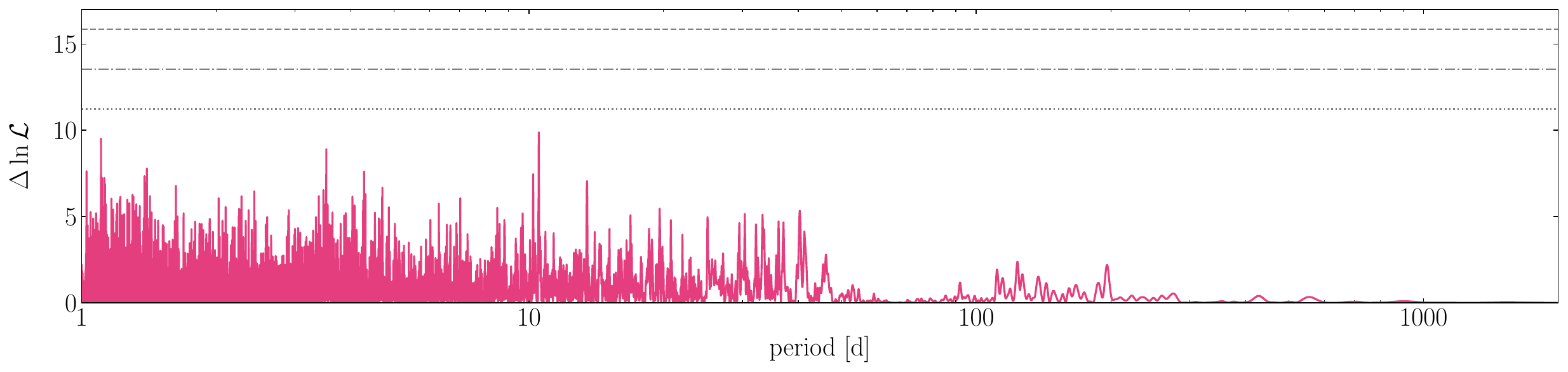}  
 \caption{RVs, MLP periodogram, and best-fit model of GJ\,581.
 {\em Top panel}: Precise Doppler measurements of GJ\,581 from HIRES (purple circles), HARPS (orange diamonds), and CARMENES (pink triangles) modeled with a three-planet self-consistent N-body model simultaneously constructed with a GP component that fits the stellar activity periodicity in the RVs. The middle subpanel shows the residuals 
 of the N-body model, where are effectively fit with the rotational kernel GP model.
The bottom subpanel shows the data residuals to the total N-body+GP model.
{\em Middle panels}: Phase folded data and N-body model for the three planetary signals as indicated in the subplots. 
{\em Bottom panel}: MLP periodogram of the RV model and GP residuals. No significant periodicity remains after the model is subtracted.
}
\label{new_data} 
\end{figure*}

\section{Orbital analysis and results}
\label{sec:OrbDynAnalysis}

\subsection{Orbital analysis tools}
\label{sec:tools2}

We used the {\tt Exo-Striker} tool for orbital parameter analysis, 
which offers orbital modeling of RV data with multi-Keplerian and with self-consistent N-body models that account for gravitational interactions between the planets. As a proxy for stellar activity, the {\tt Exo-Striker} relies on GP regression models using the {\tt celerite} package  \citep{Mackey2017}.  It provides efficient best-fit optimization algorithms and parameter sampling techniques, such as Markov chain Monte Carlo (MCMC) sampling through the {\tt emcee} sampler \citep{emcee} and nested sampling (NS) via the {\tt dynesty} sampler \citep{Speagle2020}. Numerical stability analysis of the multiple-planet configuration was performed using a customized version of the {\tt SWIFT} N-body package \citep{Duncan1998}, which also is conveniently incorporated in the {\tt Exo-Striker}.
 
We started with a multiple-planet Keplerian model, using its best-fit parameters as an initial guess for the more complex self-consistent N-body model, which we ultimately employed in our study of the GJ\,581 system. In our modeling approach, we fit the RV 
offsets and RV jitter parameters of the HARPS, HIRES, and CARMENES data sets, introducing six free parameters. The fitting also 
encompassed the spectroscopic Keplerian parameters of each planet, 
including the RV semi-amplitude $K$, orbital period $P$, 
eccentricity $e$, argument of periastron $\omega$, and the mean 
anomaly $MA$ for the epoch of the first RV data point of HIRES.
For our N-body model, the fitting parameters remained the same and in our final and most intricate orbital N-body model, we incorporated coplanar planetary inclinations $i$-tied to the same value for all planets. For the GP regression model, we adopted the rotational GP  kernel, which serves as a proxy for 
quasi-periodic stellar rotation modulation. The form of the kernel 
 as formulated by \citet{Mackey2017} is:
 
\begin{equation}
 k(\tau) = \frac{B}{2+C}e^{-\tau/L}  \left[ \cos\bigg( \frac{2\pi\tau}{P_{\rm rot}} \bigg) + (1+C) \right] ,
\label{eq:hill1}
\end{equation}

\noindent
where $P_{\rm rot}$ serves as a proxy for the stellar rotation period, $L$ represents the coherence time scale (such as the lifetime of stellar spots), $\tau$ denotes the time-lag between two data points, and $C$ acts as a balancing parameter between the periodic and nonperiodic components of the GP kernel.

The initial fitting employed a maximum likelihood 
estimator (MLE) scheme, optimizing orbital parameters through the 
Nelder-Mead simplex algorithm \citep[][]{NelderMead}. The Nelder-Mead best-fit solution served as the foundation for obtaining the 
most accurate solution, providing a robust initial guess for more 
detailed prior and posterior estimates. 
We used the posterior distribution obtained through NS to estimate parameter uncertainties and performed model comparison 
through statistical metrics such as $\Delta
\ln{\mathcal{L}}$, Bayesian information criterion (BIC), and $\Delta\ln{\mathcal{Z}}$. The 1-$\sigma$ parameter uncertainties in this work were determined as the 68.3\% 
confidence levels of the posterior parameter distribution.

\subsection{Model comparison results}
\label{sec:Model comparison results}
 
In our analysis, we assumed that the signal attributed to the hypothesized 
GJ\,581\,d planet arises from stellar activity and thus we modeled it with a GP. Consequently, our modeling 
approach incorporates a maximum of three planets for GJ\,581. We began 
with no planets and progressively included configurations with varying 
numbers of planets, assessing their statistical properties.
We compared these models by computing the difference of the $\ln{\mathcal{L}}$ values of their respective MLE fits, that is, $\Delta\ln{\mathcal{L}} = |\ln{\mathcal{L}}_{\rm complex~model}| - |\ln{\mathcal{L}}_{\rm simpler~model}|$. Further, we computed the BIC for each of our model configurations and then also calculated the differences,  $\Delta$BIC. 
The BIC depends on $\ln{\mathcal{L}}$ through 
$\mathrm{BIC} =k\ln{n}-2\ln{{\mathcal {L}}}$, where $k$ corresponds to the number of free parameters and $n$ to the number of data points used for the analysis. 
Lastly, we performed an NS analysis for which we computed the $\ln{\mathcal{Z}}$ value of the posteriors, also known as the Bayesian evidence or marginal likelihood, which comes directly from the converged posteriors obtained with {\sc dynesty}.
To identify the best model, we determined whether the above statistical indicators showed a strong preference. For $\Delta\ln{\mathcal{L}}$, we based the threshold for significant detection on the values of \citet{Anglada-Escude2016}, who proposed that $\Delta\ln{\mathcal{L}}>7$ should be considered statistically significant, while for the BIC we adopted the value of $\Delta\mathrm{BIC}>10$  as statistically significant \citep{Kass1995}. Finally, following \citet{trotta2008bayes}, we adopted a value of $\Delta\ln{\mathcal{Z}}=2$ as moderate evidence for a better-suited model, while $\Delta\ln{\mathcal{Z}}>5$ shows strong evidence for a better-suited model.

 \begin{figure*}
\includegraphics[width=9cm]{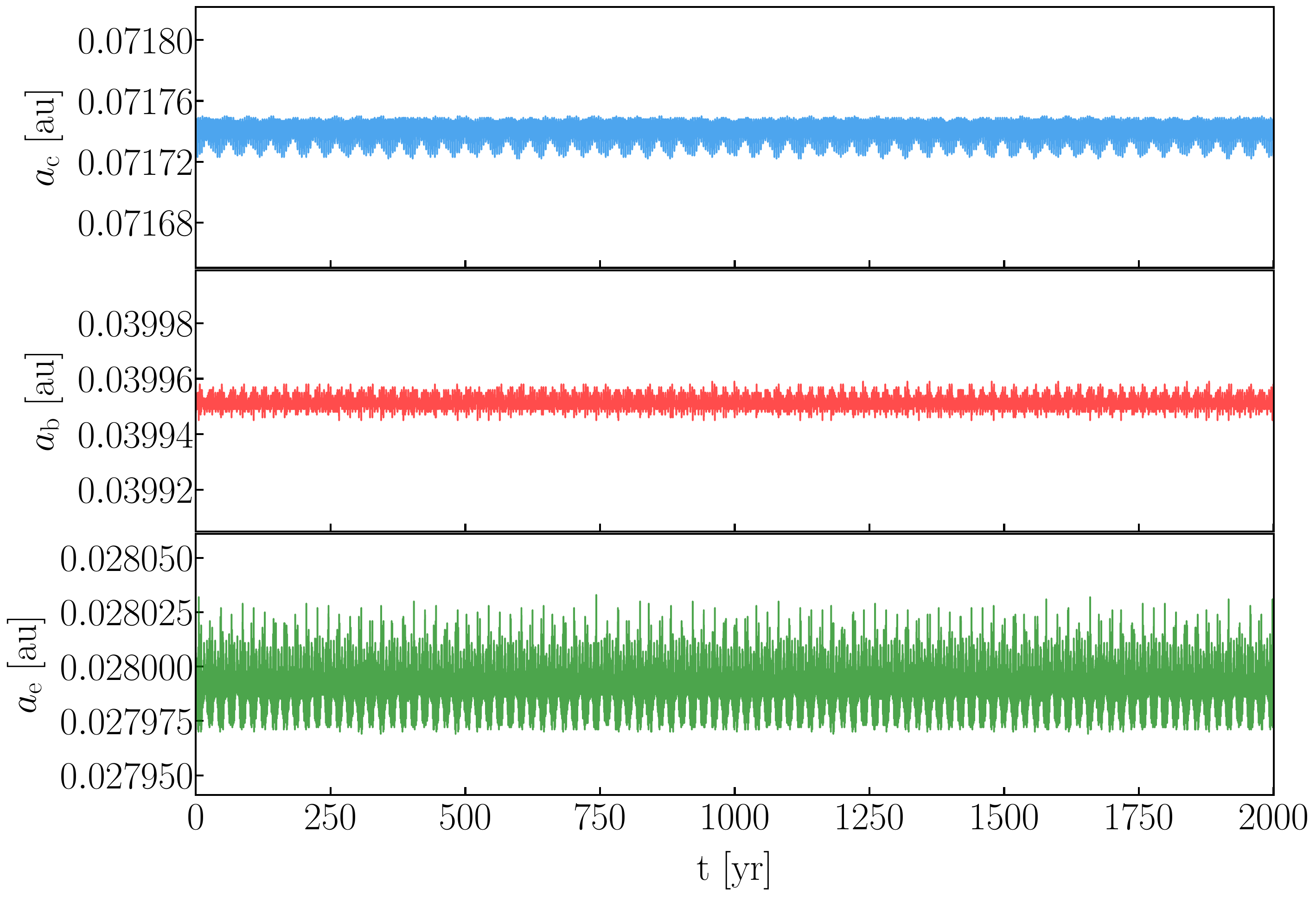}
\includegraphics[width=9cm]{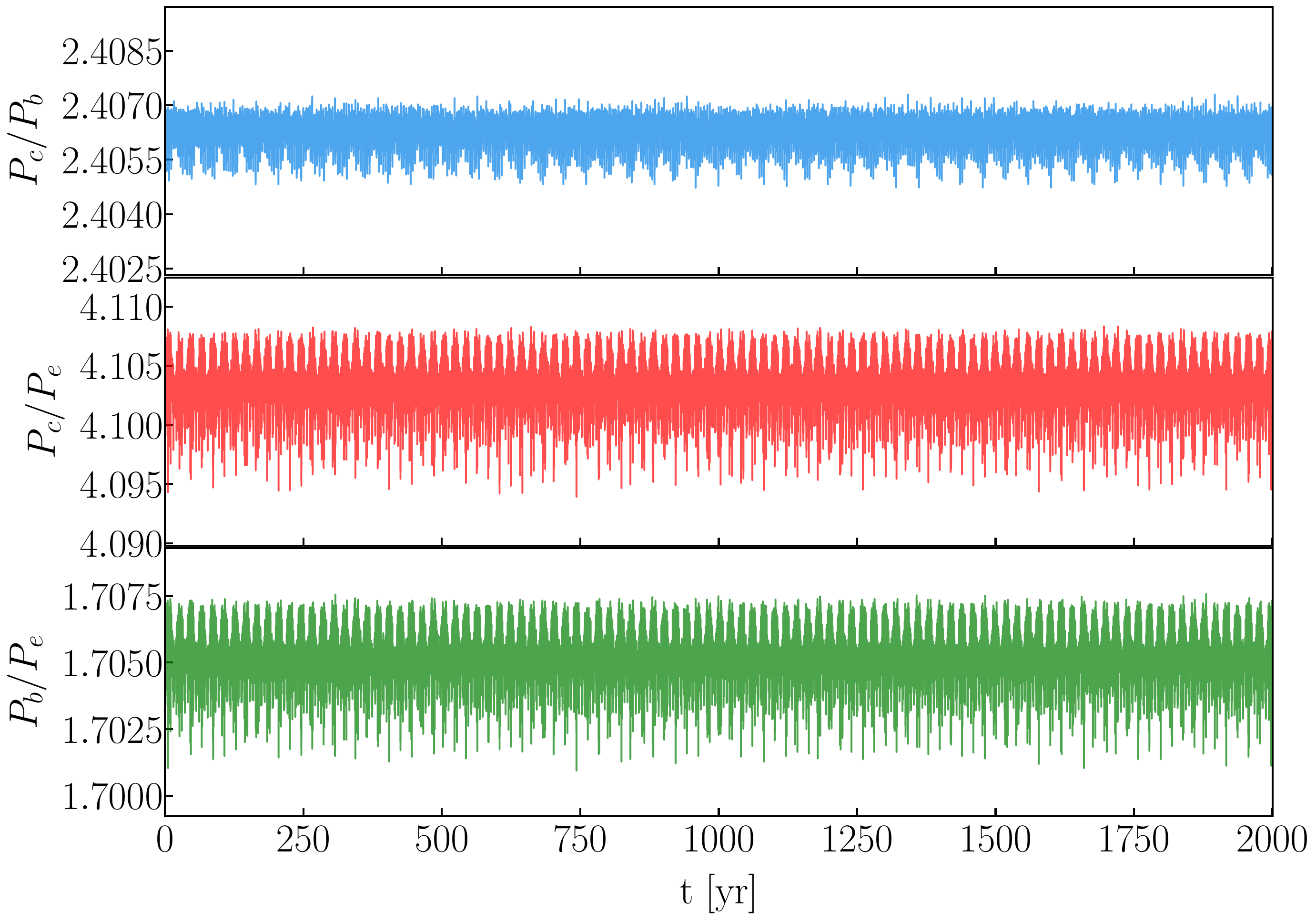}\\
\includegraphics[width=9cm]{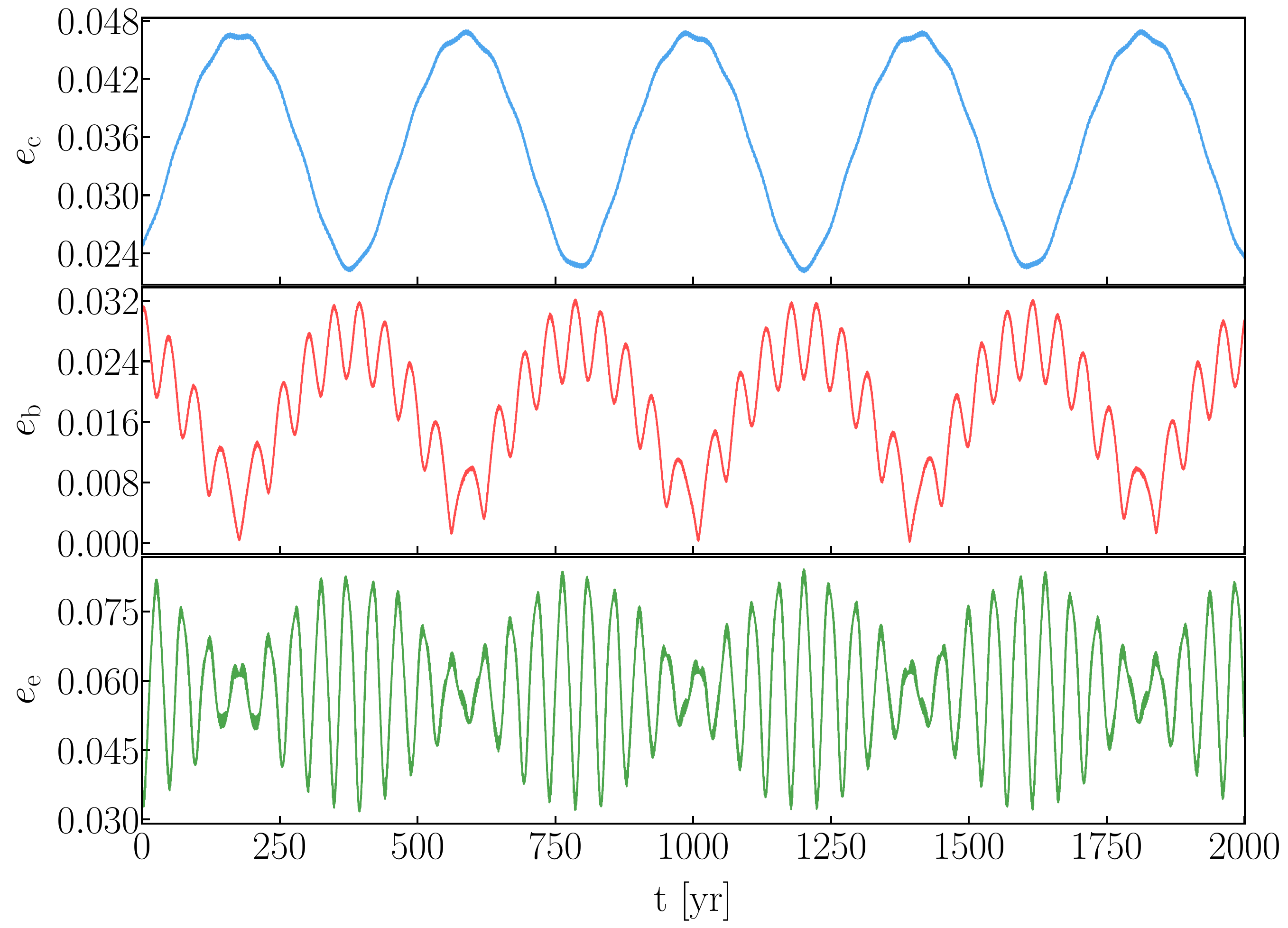}
\includegraphics[width=9cm]{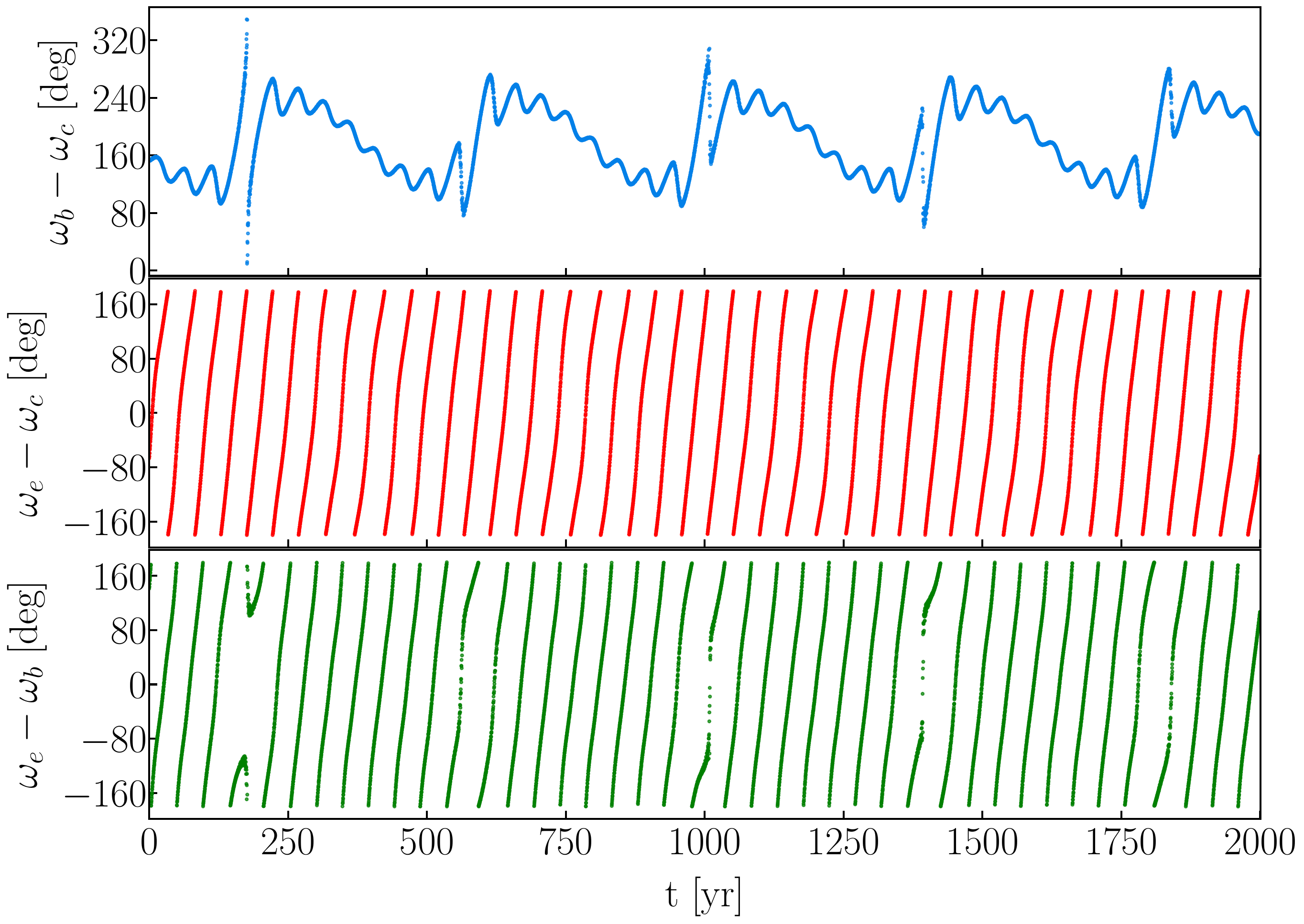}
\caption{N-body orbital evolution of the GJ\,581 system using initial orbital parameters from the median NS stable sample to represent the system's dynamical behavior for the first 2,000 yr of evolution. The top panels show the evolution of planetary semi-major axes on the left and mutual orbital period ratios are shown on the right. The bottom panels show the evolution of planetary eccentricities on the left, along with the evolution of the mutual apsidal alignment angles. No secular or MMR dynamical architecture was observed for GJ\,581.
}
\label{orb_evol}
\end{figure*}

The results from our model comparison test are given in \cref{table:stat_param}. 
First, we constructed our flat model 0, which has only the RV offsets and jitters for each data set, that is, a total number of six free parameters.
We then created models that take into account the confirmed planets by successively incorporating full Keplerian models (see \cref{sec:tools2}) using as an initial guess the RV semi-amplitude, the period, and RV signal phase as constrained by our periodogram analysis (see \cref{sec:RV Periodogram}). 
Therefore, in order, we incorporated GJ\,581\,b, with a period of 5.36\,d, to create model 1, then GJ\,581\,c, with a period of 12.9\,d, for model 2, and lastly GJ\,581\,e, with a period of 3.15\,d, for 
model 3. Model 4 contains a three-planet Keplerian model along with a GP model to capture the stellar activity, introducing more free parameters.
We used model 4 as a baseline for model 5, where we employed a self-consistent N-body dynamical model instead of a multi-Keplerian model. This approach better represents the physical properties of the system by accounting for gravitational interactions among all planets.

From model 0 to 3, all three of the quality assessment values  increased significantly.
This outcome was expected as all three planets are confirmed and fitting their signals should improve the statistical significance of the model. 
The largest improvement is predictably seen from the flat model to the first model. 
This improvement can be attributed to the strong signal GJ\,581\,b exhibits, as also seen in \cref{fig:MLP_Rv_periodograms}.
Incorporating the GP-model component in the Keplerian three-planet model led to a further significant enhancement.
Transitioning from model 4 to the N-body model 5 resulted in a 
marginal increase in both likelihood and BIC.
Interestingly, $\ln{\mathcal{Z}}$, experienced a marginal 
decrease of $\Delta\ln{\mathcal{Z}}=2.8$. Despite this marginal setback,
the decision was made to proceed with the three-planet N-body model, 
as it incorporates gravitational interactions between the planets and provides a more realistic portrayal of the physical system. The results from the coplanar edge-on NS fit in terms of median posterior parameters and uncertainties, as well parameters from the best $-\ln\mathcal{L}$ MLE search values, are presented in \cref{table:NS_post}.

Finally, we conducted an additional NS fit by employing our 
best-model configuration from model 5 and introducing the inclination of the orbital plane as an additional free parameter. 
In essence, we thus maintained the system as coplanar, but scaled the masses by a 
factor of $\sin i$, thereby constructing a posterior 
probability distribution for the dynamical masses of the GJ\,581 system.
In preparation for the coplanar inclined NS run, we systematically varied the system inclination followed by a global MLE parameter optimization.  We found that fits with $i$ lower than 10 degrees typically resulted in poor convergence due to the increased planetary masses and planet-planet interactions. Therefore, for the NS run, we opted for noninformative flat priors for $i$ within the range of 10 to 90 degrees.

Our results exhibited a marginal improvement with respect to the edge-on N-body model solution with $\Delta\ln\mathcal{L}=1.5$, but penalized by $\Delta \mathrm{BIC}=-3.14$. The Bayesian evidence, however, favors the more complex model with $\Delta\ln{\mathcal{Z}}=6.12$, sufficiently large to adopt it according to \cite{trotta2008bayes}. 
The posterior distribution yields 
$i = 47_{-13}^{+15}$\,deg. 
Consequently, 
the planets could be up to $\sim30$\% more massive.

Model 6 represents the final orbital update for the GJ\,581 system adopted in this work. The posterior distribution of the NS analysis is shown in \cref{Fig:cornerplot} and \cref{table:NS_post2} presents the medians and 1$\sigma$ uncertainties derived from the NS of the posterior distribution and the best-fit solution with a maximum  $-\ln\mathcal{L}$, found via the MLE search. 
These results enabled us to determine planetary parameters such as the periods $P_{\rm e} = 
3.1481_{-0.0004}^{+0.0004}$\,d, $P_{\rm b} = 5.3686_{-0.0001}^{+0.0001}$\,d, and $P_{\rm c} = 
12.9211_{-0.0007}^{+0.0008}$\,d, 
the eccentricities $e_{\rm e} = 0.012_{-0.008}^{+0.015}$, $e_{\rm b} = 0.0342_{-0.010}^{+0.009}$, $e_{\rm c} = 
0.032_{-0.027}^{+0.021}$, and the dynamical masses $m_{\rm e} = 2.48_{-0.42}^{+0.70} M_{\oplus}$, $m_{\rm b} = 20.50_{-3.47}^{+6.18} M_{\oplus}$, and $m_{\rm c} = 6.81_{-1.16}^{+0.21} M_{\oplus}$.
 
The best-fit model 6 is shown in the top panel of \cref{new_data}, where all three RV time series 
used in the analysis (CARMENES, HIRES, HARPS) are shown together with the respective RV residuals below. The bottom panel illustrates an MLP power spectrum of the residuals, which exhibits no significant signals.

 \subsection{Dynamical analysis}
 \label{sec:DynamicalAnalysis}

The GJ\,581 planets form a compact multi-planetary system, 
necessitating a thorough investigation of both its long-term 
stability and dynamical architecture. To address this, we employed 
the Wisdom–Holman N-body algorithm \citep[also known as MVS;][]
{Wisdom1991} with a small time step of 0.03\,d to effectively capture the evolution of the shortest-period planet.
Numerical simulations were performed for our adopted best-fit 
model 6. Commencing at the first RV epoch BJD = 2451409.762,
we conducted an integration of 10\,Myr, which covered just over one billion completed orbits of GJ\,581\,e. 
For posterior analysis, we randomly chose 1,000 configurations from the NS and integrated them for a much shorter time of a maximum of 10,000\,yr (i.e., about a million completed orbits of GJ\,581\,e). This numerical 
setup is sufficient for studying the long-term stability of the system, allowing us to construct posteriors of its dynamical parameters. 

We assessed the stability of the system by inspecting the evolution of the semi-major axis and eccentricity over time for each integrated sample. If the variations in the semi-major axis of a planet became too large or the orbital eccentricity became too large, the orbits would interfere with each other over time, indicating an unstable system. We assumed that the system would be stable if the semi-major axes remained within 5\% of the best-fit values from which the system was integrated and if the eccentricities of all of the planets remained below 0.5.  

For our adopted dynamical model and integrated posteriors we found mean period ratios of $P_{\rm b} / P_{\rm e} \sim 1.70$, $P_{\rm c} / P_{\rm b} \sim  2.41$, and $P_{\rm c} / P_{\rm e} \sim  4.10$. None of these three values strongly indicates a low-order mean motion resonance (MMR), though $P_{\rm c}$/$P_{\rm e}$  is the closest candidate to a 4:1\,MMR. 
To test if an MMR is involved in the dynamics of the GJ\,581 system, we calculated the evolution of the associated MMR angles. We tested a large variety of low- and high-order MMRs. For an $n'$:$n$ resonance between an inner planet ${\rm i}$ and an outer planet ${\rm o}$, the resonance angles are given by:
\begin{equation}
    \phi_{mnn'} = n\lambda_{\rm i} - n'\lambda_{\rm o} + (m-n)\varpi_{\rm i} - (m-n')\varpi_{\rm o},
    \label{eq:resonance_angles}
\end{equation}
where $\lambda = MA + \varpi $ is the mean longitude with $MA$ being the mean anomaly, whereas $\varpi = \omega + \Omega$ is the longitude of periastron with $\omega$ being the argument of periastron, and $\Omega$ being the longitude of the ascending node (which is undefined, thus set to $\Omega= 0$\,deg). The integer $m$ satisfies $n \leq m \leq n'$ \citep{Mardling2013}. We found no libration of any resonance angle around a fixed point. Therefore, we ruled out an MMR librating configuration of GJ\,581 based on the available RV data and our NS analysis. 

\Cref{Fig:dyn_Ns} shows the posterior probability distributions of the mean planetary eccentricities and their semi-amplitudes. All examined samples are stable for 10,000\,yr, showing rather small mean eccentricities with similarly small amplitude fluctuations around their mean values. \Cref{orb_evol} shows a snapshot of the dynamical simulation carried out for the median parameters of the coplanar-inclined NS stable samples as a representative of the overall orbital dynamics of the GJ\,581 system.
The top left panel shows the evolution of the semi-major axes of the GJ\,581 planets,  which do not vary significantly over time, suggesting a stable constellation.
When inspected more closely, the semi-major axes do vary on shorter time periods, but this variation does not affect the stability of the system. 
Similarly, the period ratios between the planets are shown in the top right subpanels, which also change slightly over the integrated time. 
The bottom panels of \cref{orb_evol} show the evolution of the orbital eccentricities, which osculate with small amplitudes with secular time scales of $\sim$\,49\,yr and $\sim$\,417\,yr, but overall the orbital geometry remains nearly circular.
We found no libration in the secular apsidal arguments $\Delta\varpi$ between the planets.
We repeated the N-body integration accounting for general relativistic precession effects \citep[MVS-GR, see][]{Trifonov2020}, but found no significant deviations in the semi-amplitude and eccentricity evolution over the investigated 5,000\,yr period and, therefore, we treated them as negligible. 
We concluded that the  GJ\,581 system maintains long-term stability without any observed MMR configurations, with the difference in the orbital arguments of periastron $\Delta\omega=\omega_{\mathrm{c}}-\omega_{\mathrm{b}}$ circulating throughout the 0\,deg to 360\,deg range.

\section{Summary}
\label{sec:ConclusionDiscussion}

We present an updated orbital analysis of the GJ\,581 system, which 
utilizes CARMENES spectroscopic data in addition to the previously 
published archival HARPS and HIRES measurements.
With these extensive data and by thorough periodogram and activity analyses, we were able to confirm that, despite its strength and periodicity, the much disputed 66.7\,d signal originates from stellar activity. This result supports the conclusions of the most recent published research, that is, the system comprises only three planets with periods of $\sim$ 3.15\,d, 5.37\,d, and 12.9\,d.
We were not able to detect the elusive stellar activity signal in the CARMENES data, possibly due to its redder wavelength coverage compared to the HIRES and HARPS spectrographs, but also and most likely because of the smaller number of measurements. 
The RV signals of GJ\,581\,b and c are significantly detected in the CARMENES data and the signal for GJ\,581\,e is strengthened when all three data sets are used together. 
We studied MLP periodograms to analyze and isolate the HARPS activity indicators and found significant signals in the H$\alpha$ index time series. 
Additionally, we found a correlation of the HARPS H$\alpha$ time series with the RV residuals. 
A subsequent analysis of the signal in both the HIRES and HARPS data with S-BGLS periodograms and coherence testing revealed robust evidence that the signal near 66.7\,d is not of planetary nature.

We conducted nested sampling analyses on various multiple-planet models to the available RV data, 
encompassing both Keplerian and dynamical models featuring a GP kernel. 
The aim was to refine and constrain the parameters of the three confirmed 
remaining planets GJ\,581\,e, GJ\,581\,b, and GJ\,581\, and to effectively filter 
the RV signal induced by stellar activity. Our findings indicate that a dynamical model, when coupled with a GP model component, provides a significant improvement 
representation of the RV data compared to a nonperturbed multi-Keplerian model. 
Our results suggest that the RV data cover a sufficiently long baseline to detect secular dynamical perturbations in the GJ\,581 system. 
Additionally, we ran an N-body model scheme that includes the line-of-sight inclination as an additional fitting parameter to constrain the dynamical masses of the planets.
The latter model converged on an inclination of $i = 47.0_{-13.0}^{+14.6}$ degrees, yielding dynamical 
planet masses of $2.48_{-0.42}^{+0.70}$ $M_{\oplus}$, $20.50_{-3.47}^{+6.18}$ $M_{\oplus}$, and $6.81_{-1.16}^{+0.21}$ $M_{\oplus}$ for GJ\,581\,e, GJ\,581\,b, and GJ\,581\,c, respectively.
These masses and the model are adopted as our final orbital and mass estimates for the 
GJ\,581 system, indicating that the planets are approximately 30\% more massive than 
their previously reported masses, which assumed a coplanar system, and therefore, minimum masses.

We performed a comprehensive long-term stability analysis on the coplanar inclined system posteriors. Our findings indicate that the GJ\,581 system demonstrates long-term stability, with no evidence of MMR dynamics. The system is characterized by low-eccentricity oscillations, showcasing two dominant secular time scales of $\sim$49 and $\sim$417 years.

The nearby M dwarf star GJ\,581 is one of the most important laboratories to study exoplanet dynamics, evolution, and planet formation in general. Our work reveals the most complete dynamical architecture of the GJ\,581 planetary system extracted from Doppler data. The new insight into the orbital and dynamical configuration of
GJ\,581 provides a new focus on probing the planetary formation and evolution processes in densely populated multi-planet systems around low-mass stars.



\begin{acknowledgements}

 
  CARMENES is an instrument at the Centro Astron\'omico Hispano en Andaluc\'ia (CAHA) at Calar Alto (Almer\'{\i}a, Spain), operated jointly by the Junta de Andaluc\'ia and the Instituto de Astrof\'isica de Andaluc\'ia (CSIC).
  CARMENES was funded by the Max-Planck-Gesellschaft (MPG), 
  the Consejo Superior de Investigaciones Cient\'{\i}ficas (CSIC),
  the Ministerio de Econom\'ia y Competitividad (MINECO) and the European Regional Development Fund (ERDF) through projects FICTS-2011-02, ICTS-2017-07-CAHA-4, and CAHA16-CE-3978, 
  and the members of the CARMENES Consortium 
  (Max-Planck-Institut f\"ur Astronomie,
  Instituto de Astrof\'{\i}sica de Andaluc\'{\i}a,
  Landessternwarte K\"onigstuhl,
  Institut de Ci\`encies de l'Espai,
  Institut f\"ur Astrophysik G\"ottingen,
  Universidad Complutense de Madrid,
  Th\"uringer Landessternwarte Tautenburg,
  Instituto de Astrof\'{\i}sica de Canarias,
  Hamburger Sternwarte,
  Centro de Astrobiolog\'{\i}a and
  Centro Astron\'omico Hispano-Alem\'an), 
  with additional contributions by the MINECO, 
  the Deutsche Forschungsgemeinschaft (DFG) through the Major Research Instrumentation Programme and Research Unit FOR2544 ``Blue Planets around Red Stars'', 
  the Klaus Tschira Stiftung, 
  the states of Baden-W\"urttemberg and Niedersachsen, 
  and by the Junta de Andaluc\'{\i}a. 
  We acknowledge financial support from the Agencia Estatal de Investigaci\'on (AEI/10.13039/501100011033) of the Ministerio de Ciencia e Innovaci\'on and the ERDF ``A way of making Europe'' through projects 
  PID2022-137241NB-C4[1:4],	
  PID2021-125627OB-C31,		
  PID2019-109522GB-C5[1:4],	
  and the Centre of Excellence ``Severo Ochoa'' and ``Mar\'ia de Maeztu'' awards to the Instituto de Astrof\'isica de Canarias (CEX2019-000920-S), Instituto de Astrof\'isica de Andaluc\'ia (CEX2021-001131-S) and Institut de Ci\`encies de l'Espai (CEX2020-001058-M).
  This work was also funded by the Generalitat de Catalunya/CERCA programme, 
  the DFG through project No. KU 3625/2-1 of the DFG Research Unit FOR2544,
  the Bulgarian National Science Foundation (BNSF) program ``VIHREN-2021'' project No. KP-06-DV/5, 
  and NASA under award number NNA16BD14C for NASA Academic Mission Services.


We thank the anonymous referee for the excellent comments that helped to improve the quality of this work.
\end{acknowledgements}

\bibliographystyle{aa} 
\bibliography{GJ581} 


\begin{appendix} 

\section{Additional figures and tables}

\label{appendix}

\begin{table}[htbp!]
\begin{minipage}{\textwidth}
\centering
\caption{{The adopted priors of the Keplerian and N-body model NS runs. Their meanings are $\mathcal{U}$ -- Uniform (flat) prior, $\mathcal{N}$ -- Gaussian prior,  and $\mathcal{J}$ -- Jeffrey's (log-uniform) priors.}}
\centering
\label{table:priors}

\begin{tabular}{lrrrrrrrr}     

\hline\hline  \noalign{\vskip 0.7mm}
Parameter \hspace{10.0 mm}& \hspace{30.0 mm} GJ\,581\,e & \hspace{30.0 mm} GJ\,581\,b & \hspace{30.0 mm} GJ\,581\,c \\
\hline \noalign{\vskip 0.7mm}

    $K$ [m\,s$^{-1}$]             & $\mathcal{U}$(     1.0,     3.0)& $\mathcal{U}$(     8.0,    16.0)& $\mathcal{U}$(     1.0,     5.0)\\
    $P$ [day]                     & $\mathcal{U}$(     3.1,     3.2)& $\mathcal{U}$(     5.3,     5.4)& $\mathcal{U}$(    12.8,    13.0)\\
    $e$                           & $\mathcal{U}$(     0.0,     0.2)& $\mathcal{U}$(     0.0,     0.1)& $\mathcal{U}$(     0.0,     0.2)\\
    $\omega$ [deg]                & $\mathcal{U}$(     0.0,   360.0)& $\mathcal{U}$(     0.0,   360.0)& $\mathcal{U}$(     0.0,   360.0)\\
    $M_{\rm 0}$ [deg]             & $\mathcal{U}$(  $-$180.0,   360.0)& $\mathcal{U}$(     0.0,   360.0)& $\mathcal{U}$($-$180.0,   360.0)\\
    $i$ [deg]                     & &$\mathcal{U}$(     10.0,    90.0)& & \\
    RV$_{\rm off}$ CARM. [m\,s$^{-1}$]& &$\mathcal{U}$(    $-$5.0,     5.0)\\
    RV$_{\rm off}$ HIRES [m\,s$^{-1}$]& & $\mathcal{U}$(    $-$5.0,     5.0)\\
    RV$_{\rm off}$ HARPS [m\,s$^{-1}$]& &$\mathcal{U}$(    $-$5.0,     5.0)\\
    RV$_{\rm jit}$ CARM. [m\,s$^{-1}$]& &$\mathcal{J}$(     0.01,     5.0)\\
    RV$_{\rm jit}$ HIRES [m\,s$^{-1}$]& &$\mathcal{J}$(     0.01,     5.0)\\
    RV$_{\rm jit}$ HARPS [m\,s$^{-1}$]& &$\mathcal{J}$(     0.01,     5.0)\\
    RV GP$_{\rm Rot.}$ Amp.       & &$\mathcal{U}$(     1.0,    20.0)\\
    RV GP$_{\rm Rot.}$ timescale  & &$\mathcal{U}$(   400.0,  6000.0)\\
    RV GP$_{\rm Rot.}$ Period     & &$\mathcal{N}$(     66.7,   10.0)\\
    RV GP$_{\rm Rot.}$ fact.      & &$\mathcal{U}$(     0.05,     0.5)\\
    \\
\hline \noalign{\vskip 0.7mm}
    
\end{tabular}


\end{minipage}
\end{table}

\begin{table}[htp]
\begin{minipage}{\textwidth}
\caption{HARPS Doppler measurements and activity index measurements of GJ 581 evaluated using SERVAL. First 8 Data points. Full version available at the CDS} 
\label{table:HARPS_1} 

\centering  
\resizebox{\textwidth}{!}{
\begin{tabular}{ccccccccccccccccccc} 

\hline\hline    
\noalign{\vskip 0.5mm}

Epoch [JD] & RV [m\,s$^{-1}$] & $\sigma_{\rm RV}$ [m\,s$^{-1}$]  &  H$\alpha$ & $\sigma_{\rm H\alpha}$  & Na\,D$_1$ & $\sigma_{\rm Na\,{D_1}}$ & Na\,D$_2$ & $\sigma_{\rm Na\,{D_2}}$   &  CRX & $\sigma_{\rm CRX}$   &  dLW & $\sigma_{\rm dLW}$   &  FWHM & $\sigma_{\rm FWHM}$  & CON & $\sigma_{\rm CON}$  &   BIS & $\sigma_{\rm BIS}$ \\  

\hline     
\noalign{\vskip 0.5mm}    

2453152.713   &   $-$11.637   &    0.939 &   0.946   &    0.001 &   0.130   &    0.002 &   0.163   &    0.002 &   5.743   &    5.072 &    $-$4.886   &     1.818   &    3000.181 &    45.003   &  24.838   &    0.373 &    $-$11.619   &     $-$0.174   \\ 
2453158.663   &   $-$19.979   &    1.194 &   0.946   &    0.002 &   0.146   &    0.002 &   0.180   &    0.003 &   20.991   &    11.118 &    18.438   &     2.003   &    2966.956 &    44.504   &  23.977   &    0.360 &    $-$11.038   &     $-$0.166   \\ 
2453520.745   &   10.063   &    1.201 &   0.942   &    0.002 &   0.156   &    0.003 &   0.184   &    0.003 &   $-$5.003   &    8.719 &    $-$2.392   &     1.525   &    2995.010 &    44.925   &  24.658   &    0.370 &    $-$10.463   &     $-$0.157   \\ 
2453573.512   &   1.502   &    1.216 &   0.932   &    0.002 &   0.140   &    0.003 &   0.172   &    0.003 &   $-$11.034   &    7.562 &    4.946   &     1.367   &    2996.826 &    44.952   &  24.705   &    0.371 &    $-$8.543   &     $-$0.128   \\ 
2453574.522   &   9.343   &    1.003 &   0.936   &    0.002 &   0.136   &    0.002 &   0.169   &    0.003 &   2.030   &    5.088 &    7.272   &     1.358   &    3004.475 &    45.067   &  24.707   &    0.371 &    $-$10.351   &     $-$0.155   \\ 
2453575.481   &   2.660   &    0.860 &   0.932   &    0.001 &   0.148   &    0.002 &   0.178   &    0.002 &   23.286   &    5.455 &    $-$17.185   &     1.341   &    2997.273 &    44.959   &  25.023   &    0.375 &    $-$9.780   &     $-$0.147   \\ 
2453576.536   &   $-$7.076   &    1.003 &   0.940   &    0.001 &   0.139   &    0.002 &   0.170   &    0.002 &   4.131   &    5.958 &    $-$1.318   &     1.210   &    3002.883 &    45.043   &  24.828   &    0.372 &    $-$6.175   &     $-$0.093   \\ 
2453577.593   &   $-$10.592   &    1.170 &   0.944   &    0.002 &   0.137   &    0.002 &   0.166   &    0.003 &   $-$0.569   &    6.356 &    4.872   &     1.196   &    3002.757 &    45.041   &  24.685   &    0.370 &    $-$3.508   &     $-$0.053   \\  
  
\hline           
\end{tabular}}

\end{minipage}
\end{table}
\begin{table}[htp]
\begin{minipage}{\textwidth}
\caption{CARMENES RV measurements and activity index measurements of GJ 581 evaluated by SERVAL and their corresponding uncertainties. First 8 Data points. Full version available at the CDS} 
\label{table:CARM_MAIN} 

\centering  
\resizebox{\textwidth}{!}{
\begin{tabular}{cccccccccccccccccc} 

\hline\hline    
\noalign{\vskip 0.5mm}

Epoch [JD] & RV [m\,s$^{-1}$] & $\sigma_{\rm RV}$ [m\,s$^{-1}$]  &  H$\alpha$ & $\sigma_{\rm H\alpha}$  & Na\,D & $\sigma_{\rm Na\,D }$   &  CRX & $\sigma_{\rm CRX}$   &  dLW & $\sigma_{\rm dLW}$   &  FWHM & $\sigma_{\rm FWHM}$  & CON & $\sigma_{\rm CON}$ &   BIS & $\sigma_{\rm BIS}$\\  

\hline     
\noalign{\vskip 0.5mm}    

2457415.763   &   $-$10.377   &    1.127 &   0.917   &    0.001 &   0.085   &    0.002 &   13.949   &    9.106 &   5.839   &    0.944 &    4395.973   &     24.397   &    22.871 &    0.089   &  $-$15.180   &    4.148    \\ 
2457418.762   &   13.277   &    1.150 &   0.923   &    0.002 &   0.097   &    0.003 &   21.726   &    9.368 &   $-$2.817   &    1.163 &    4398.735   &     23.197   &    22.915 &    0.084   &  $-$8.544   &    5.263    \\ 
2457422.738   &   $-$4.847   &    1.656 &   0.911   &    0.001 &   0.080   &    0.002 &   23.931   &    9.329 &   0.987   &    1.045 &    4399.537   &     23.337   &    22.923 &    0.085   &  $-$10.775   &    4.129    \\ 
2457466.713   &   14.703   &    1.343 &   0.903   &    0.002 &   0.090   &    0.003 &   3.858   &    9.386 &   0.580   &    1.153 &    4404.342   &     23.939   &    22.965 &    0.087   &  $-$9.000   &    4.541    \\ 
2457476.636   &   0.932   &    1.268 &   0.905   &    0.001 &   0.093   &    0.002 &   10.016   &    8.527 &   $-$0.767   &    1.063 &    4396.872   &     22.554   &    22.939 &    0.082   &  $-$10.326   &    3.805    \\ 
2457490.597   &   $-$11.618   &    1.346 &   0.904   &    0.002 &   0.087   &    0.005 &   10.489   &    11.332 &   $-$11.920   &    1.487 &    4381.150   &     24.195   &    23.091 &    0.089   &  $-$11.098   &    6.341    \\ 
2457493.601   &   14.607   &    1.297 &   0.909   &    0.002 &   0.099   &    0.005 &   12.559   &    12.321 &   $-$8.169   &    1.786 &    4391.286   &     23.004   &    23.059 &    0.084   &  $-$8.261   &    5.930    \\ 
2457503.558   &   7.256   &    1.210 &   0.917   &    0.001 &   0.086   &    0.003 &   6.142   &    8.831 &   3.457   &    1.147 &    4390.222   &     23.975   &    22.936 &    0.087   &  $-$8.837   &    3.932    \\   
  
\hline           
\end{tabular}}

\end{minipage}
\end{table}

\begin{figure*}
\centering
\includegraphics[width=0.6\textwidth]{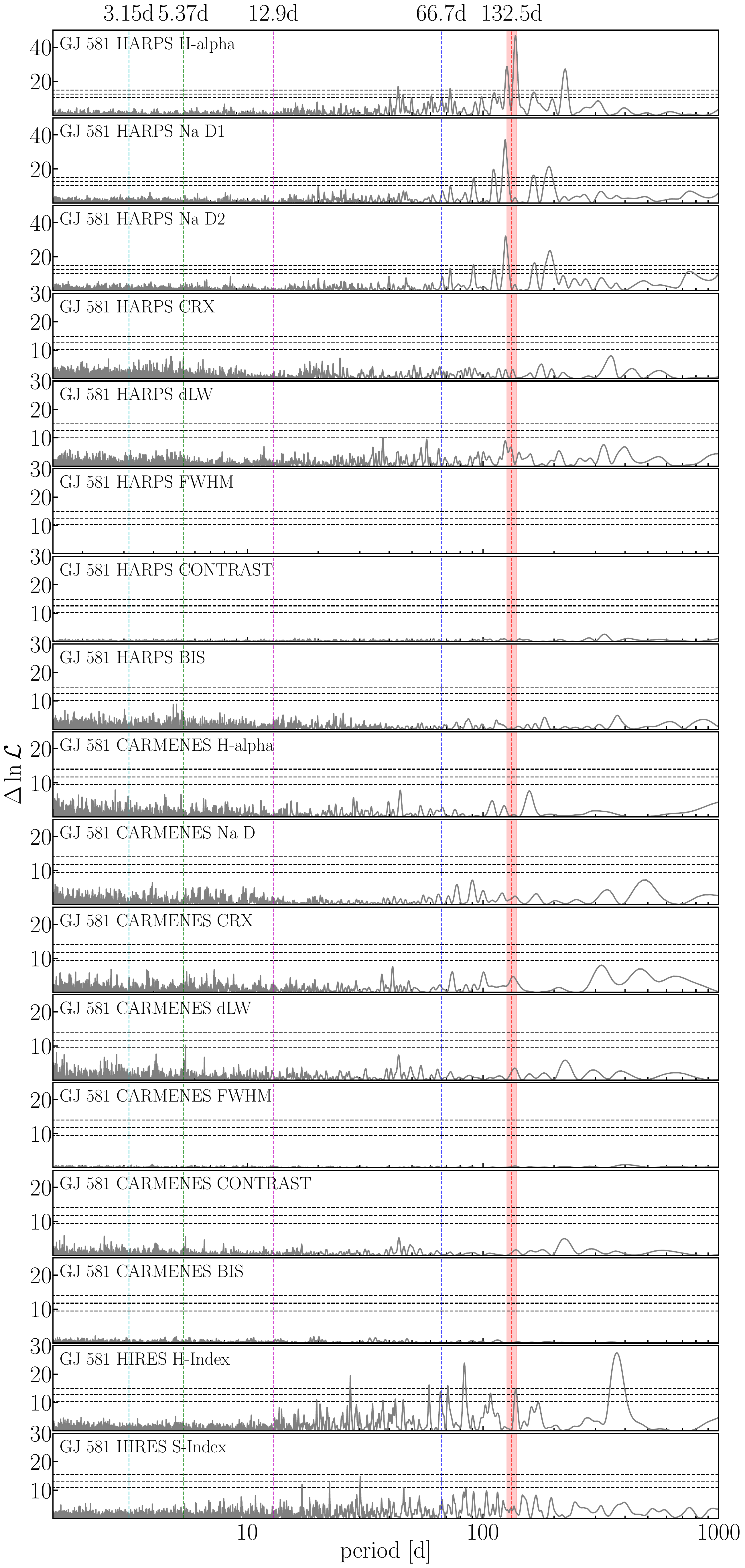} 
     \caption{Full MLP power spectrum for all RV time series used, as well as their respective activity time series. Horizontal dashed lines indicate FAP levels of 10\%, 1\%, and 0.1\%. The cyan, green, magenta, and blue vertical lines indicate the period of GJ\,581\,e, GJ\,581\,b, GJ\,581\,c and the 66.7\,d signal. respectively.
     }
   \label{Fig:Fullperiodogram}
\end{figure*}

\begin{figure*}
\includegraphics[width=\textwidth]{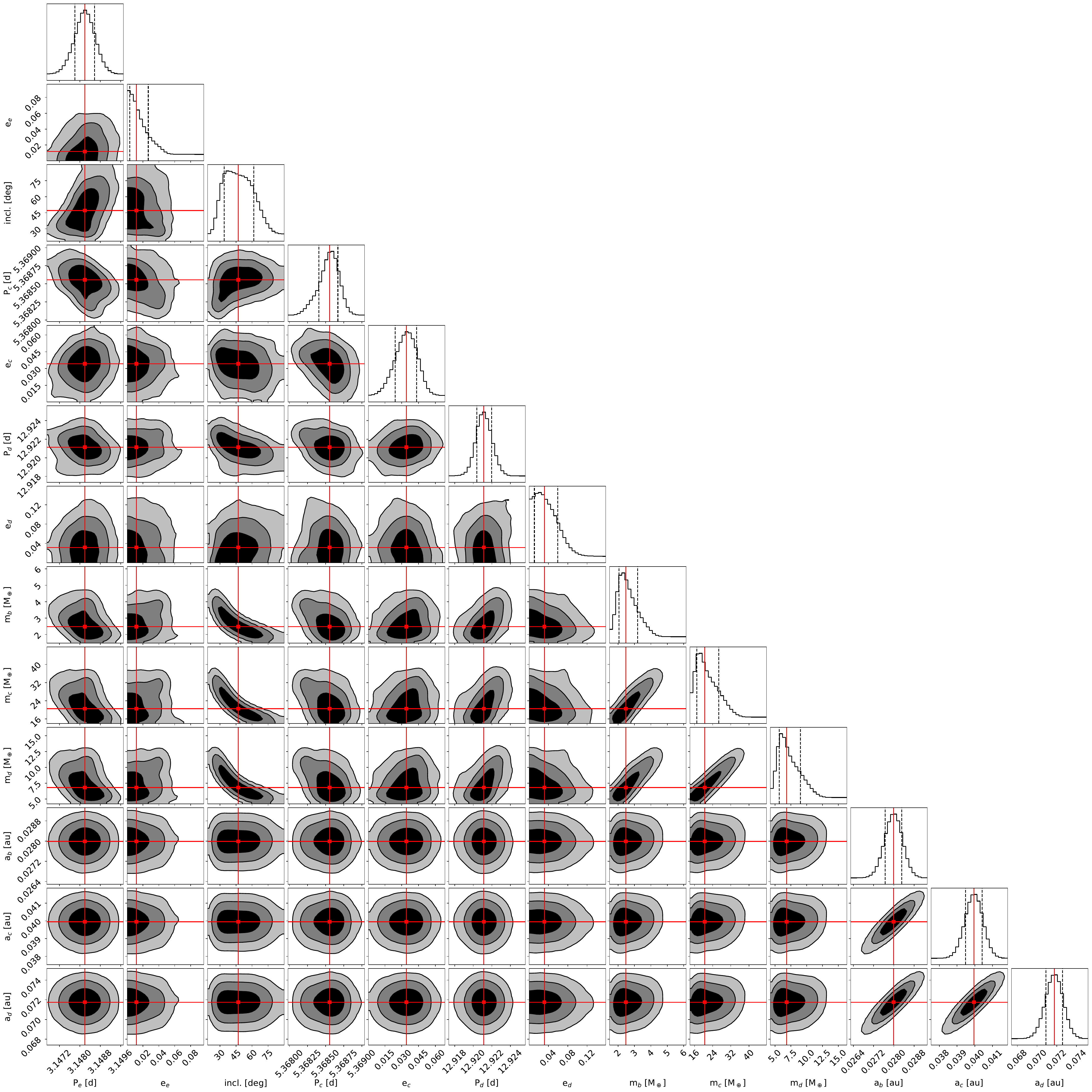}  
\caption{Distribution of the NS posteriors of a coplanar inclined configuration of the GJ\,581 system. The two-dimensional contours indicate 1-, 2-, and 3-$\sigma$ confidence levels of the posterior distribution. The red crosses indicate the median of the posterior probability distribution. Top to bottom and left to right:  mean orbital period $P_{\rm pl}$, mean planetary eccentricities $e_{\rm pl}$, planetary masses (in $M_{\oplus}$), and the semi-major axes (in au).}
\label{Fig:cornerplot}
\end{figure*} 

\begin{figure*}
\includegraphics[width=\textwidth]{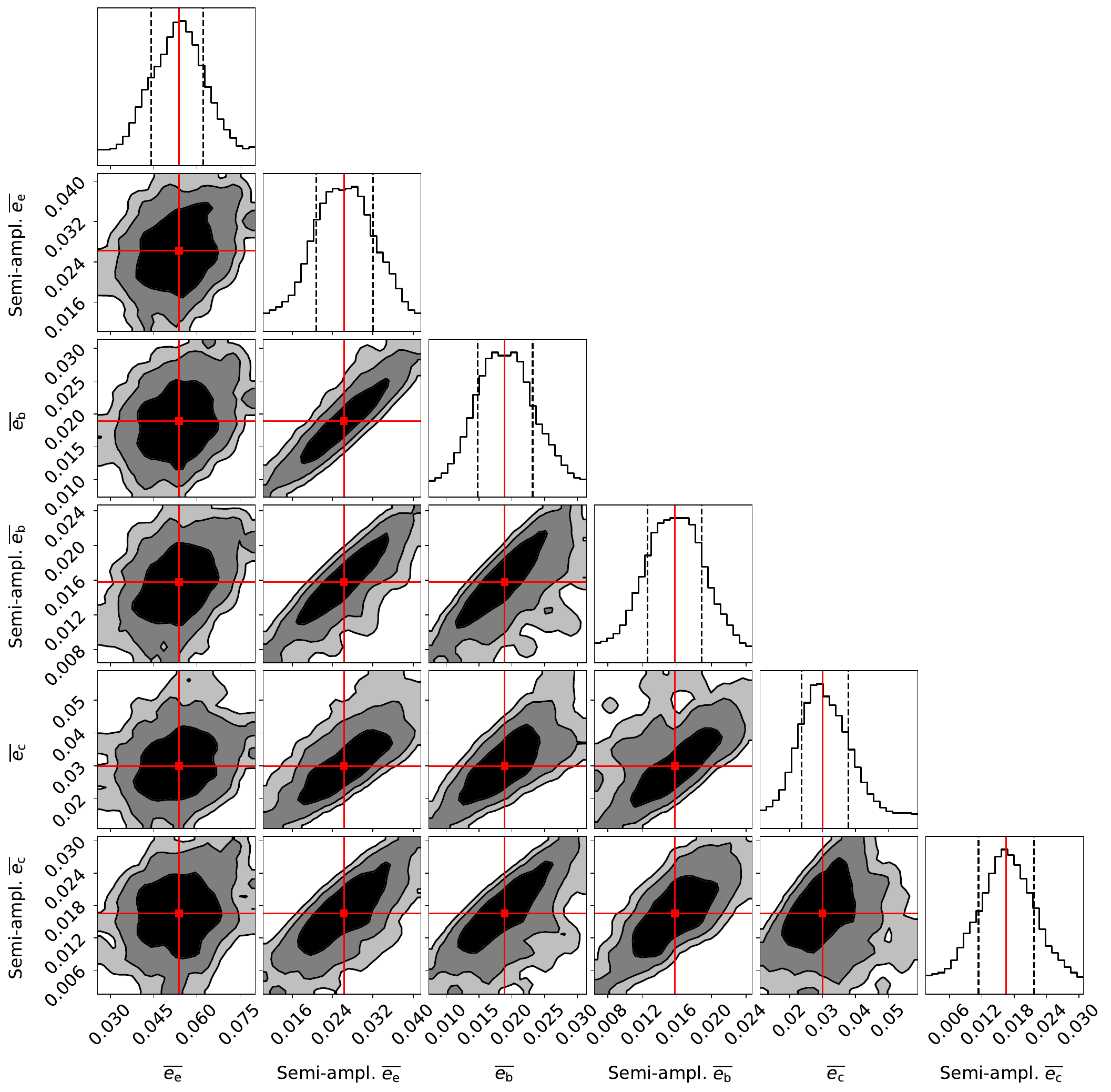}  
\caption{Distribution of 1,000 randomly sampled NS posteriors from the coplanar inclined model (see \cref{Fig:cornerplot}) integrated over 10,000 years. The two-dimensional contours represent the 1-, 2-, and 3-$\sigma$ confidence intervals of the stable posterior distribution.
From top to bottom, the corner plot shows the mean osculating planetary eccentricities and osculating eccentricity semi-amplitudes for the three planets, pivotal for assessing the GJ\,581 system's dynamical properties. 
The median values of the complete posterior distributions are indicated by red lines.
}
\label{Fig:dyn_Ns}
\end{figure*}

\begin{figure*}
    \includegraphics[width=18cm]{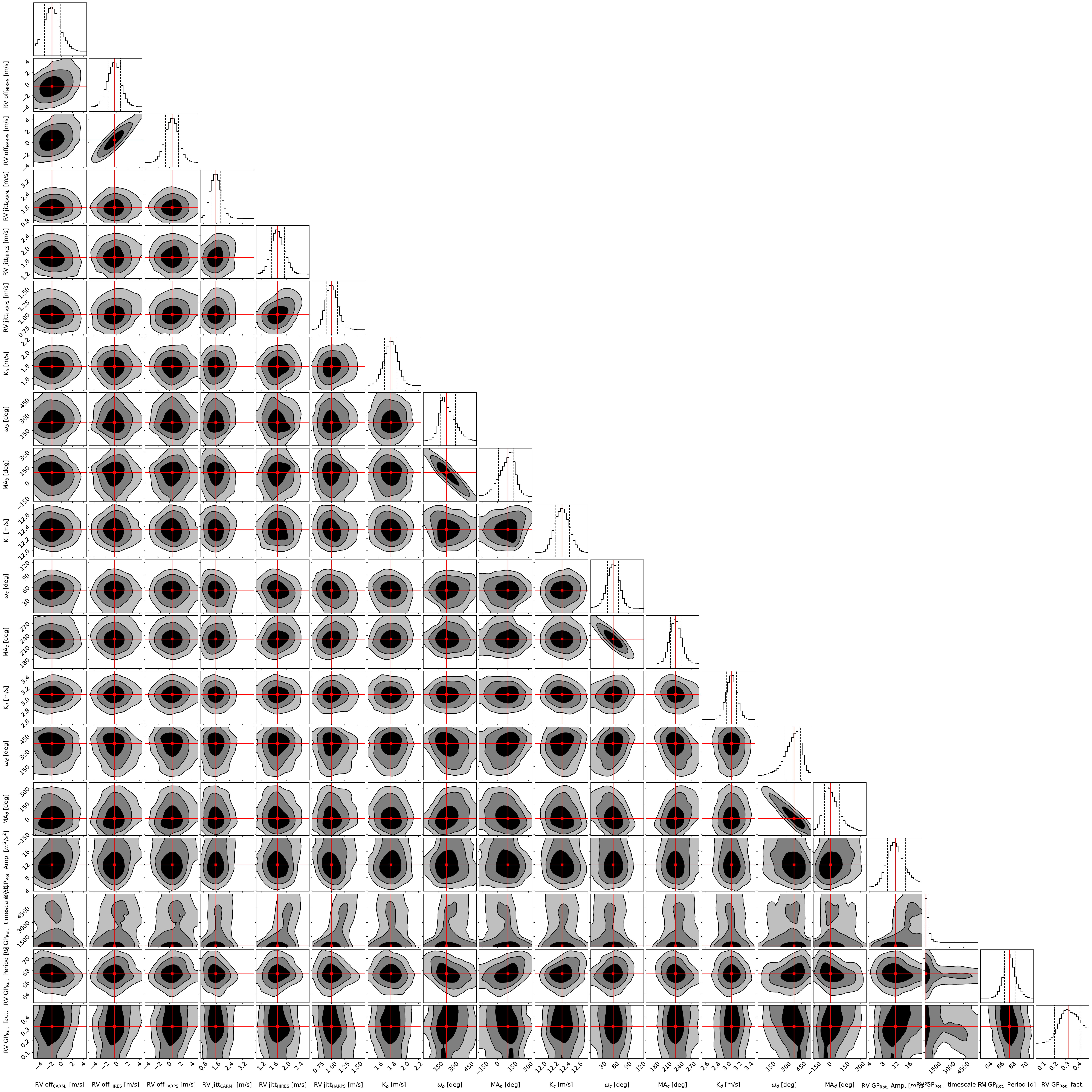}  
    \caption{The remaining posteriors of the dynamical properties of the best-fit dynamical model that are not displayed in the main text. The two-dimensional contours indicate 1-, 2- and 3-$\sigma$ confidence levels of the posterior distribution. The red crosses indicate the median of the posterior probability distribution. Top to bottom and left to right: The RV offsets of CARMENES, HIRES and HARPS, respectively, the RV jitter, as well as the mean semi-amplitude $K$, mean argument of periapsis $\omega$, the mean anomaly $MA$ of each planet and the GP rotational kernel amplitude, time scale, period, and factor. 
    }
    \label{Fig:cornerplot_App}
\end{figure*}

\end{appendix}

\end{document}